\documentclass[amsmath,amssymb,aps,prl,superscriptaddress,twocolumn,floatfix]{revtex4-1}
\usepackage{graphicx}% Include figure files
\usepackage{dcolumn}% Align table columns on decimal point
\usepackage{bm}
\usepackage[usenames]{color}
\usepackage{xcolor}
\usepackage{soul}
\usepackage{comment}

\usepackage{hyperref} 
\usepackage{physics}
\usepackage{bookmark}
\usepackage{mathtools}
\usepackage{subfigure}

\usepackage{wasysym}

\newcounter{para}
\newcommand{\para}{\par\refstepcounter{para}\textbf{{\color{cyan}[\thepara]}}\space}

\let\para\relax

\newcommand{\Heff}{H_{\textnormal{eff}}}
\begin{document}

\title{Fractionalization in Fractional Correlated Insulating States at $n\pm 1/3$ filled twisted bilayer graphene}

\author{Dan Mao}
\author{Kevin Zhang}
\affiliation{Laboratory of Atomic and Solid State Physics, Cornell University, 142 Sciences Drive, Ithaca NY 14853-2501, USA}

\author{Eun-Ah Kim}
\affiliation{Laboratory of Atomic and Solid State Physics, Cornell University, 142 Sciences Drive, Ithaca NY 14853-2501, USA}
\date{October 2021}
\affiliation{Radcliffe Institute for Advanced Study at Harvard, Harvard University, 10 Garden Street, Cambridge MA 02138, USA}
\affiliation{Department of Physics, Harvard University, 17 Oxford Street, Cambridge MA 02138, USA}
\affiliation{Department of Physics, Ewha Womans University}

\begin{abstract}
Fractionalization without time-reversal symmetry breaking is a long-sought-after goal in the study of correlated phenomena. 
The earlier proposal of correlated insulating states at $n \pm 1/3$ filling in twisted bilayer graphene and recent experimental observations of insulating states at those fillings strongly suggest that moir\'e graphene systems provide a new platform to realize time-reversal symmetric fractionalized states. However, the nature of fractional excitations and the effect of quantum fluctuation on the fractional correlated insulating states are unknown. We show that excitations of the fractional correlated insulator phases in the strong coupling limit carry fractional charges and exhibit fractonic restricted mobility. Upon introduction of quantum fluctuations, the resonance of ``lemniscate" structured operators drives the system into ``quantum lemniscate liquid (QLL)" or ``quantum lemniscate solid (QLS)".  %{\color{purple}
We find an emergent $U(1)\times U(1)$ 1-form symmetry unifies distinct motions of the fractionally charged excitations in the strong coupling limit and in the QLL phase while providing a new mechanism for fractional excitations in two-dimension.  We predict emergent Luttinger liquid behavior upon dilute doping in the strong coupling limit due to restricted mobility and discuss implications at a general  $n \pm 1/3$ filling. 
%} 

\end{abstract}

\maketitle
\para 
Fractionalization, where the quantum number of low energy excitations is a fraction of the physical constituents (such as electrons), epitomizes strong correlation effects. With reduced phase space amplifying the correlation effects, fractionalization does not require magnetic field in 1D systems \cite{Su1979Phys.Rev.Lett.,Pham2000Phys.Rev.B, Imura2002Phys.Rev.Ba,Orgad2001Phys.Rev.Lett., Steinberg2008NaturePhys}. However, in higher dimensions, fractionalization has only been confirmed with breaking of time-reversal symmetry either under fractional quantum Hall settings
 \cite{Laughlin1983Phys.Rev.Lett.,Martin2004Science}
or spontaneous time-reversal symmetry breaking in fractional Chern insulators~\cite{Regnault2011Phys.Rev.X,Xie2021Nature}.
Theoretical proposals for fractionalization without time-reversal symmetry breaking have invoked the effects of geometric frustration with local constraints, giving rise to emergent gauge theories in spin models and quantum dimer models \cite{anderson1987resonating,Rokhsar1988Phys.Rev.Lett., affleck1988large,read1991large,wen1991meanfield,moessner2001resonating,senthil2000z2,Kitaev_2003,Kitaev_2006,castelnovo2008magnetic}.
More recently, the notion of constraints has been taken to new directions with the advent of fracton models characterized by excitations with restricted mobility \cite{chamon2005quantum,haah2011local,yoshida2013exotic,Vijay2015Phys.Rev.B, vijay2016fracton,Nandkishore2019Annu.Rev.Condens.MatterPhys.,Pretko2020Int.J.Mod.Phys.A}. 
While exactly solvable models offer theoretical insight \cite{Rokhsar1988Phys.Rev.Lett., chamon2005quantum,haah2011local,yoshida2013exotic,Vijay2015Phys.Rev.B}, finding a physical realization has been challenging. 

\para
The recent observation of time-reversal invariant incompressible states (i.e., zero Chern number) at fractional filling in twisted bilayer graphene \cite{Xie2021Nature} presents a new platform for a strongly correlated state at fractional filling. While the nature of the observed states is still largely unknown, two of us predicted that 
``fidget spinner"-shaped Wannier orbitals of twisted bilayer graphene can lead to a correlated insulating phase at fractional filling due to the geometric constraints imposed by the shape of the orbitals \cite{Zhang2022}. While the extensive ground state degeneracy observed in the strong coupling limit \cite{Zhang2022} implies novel geometrical frustration effects in widely available physical platforms, little is known about the nature of excitations and effects of quantum fluctuations. 
%{\color{purple} 
In this letter, we evince the fractionalization of doped holes and fractonic nature of the fractionally charged excitations in the strong coupling limit. Furthermore, we derive a resonance in the lemniscate configuration of Wannier orbitals as the leading quantum fluctuation effect that can result in a QLL/QLS (quantum lemniscate liquid/solid) phase. We find an emergent $U(1)\times U(1)$ 1-form symmetry at low energy and relate the fracton-like behavior of the excitations to the non-trivial string operator under the 1-form symmetry. Finally, we generalize our formalism to other third fillings and twisted trilayer graphene and discuss experimental prospects of detecting the proposed fractionalization. 
%}

\begin{figure}[t]
    \includegraphics[width=.4\textwidth]{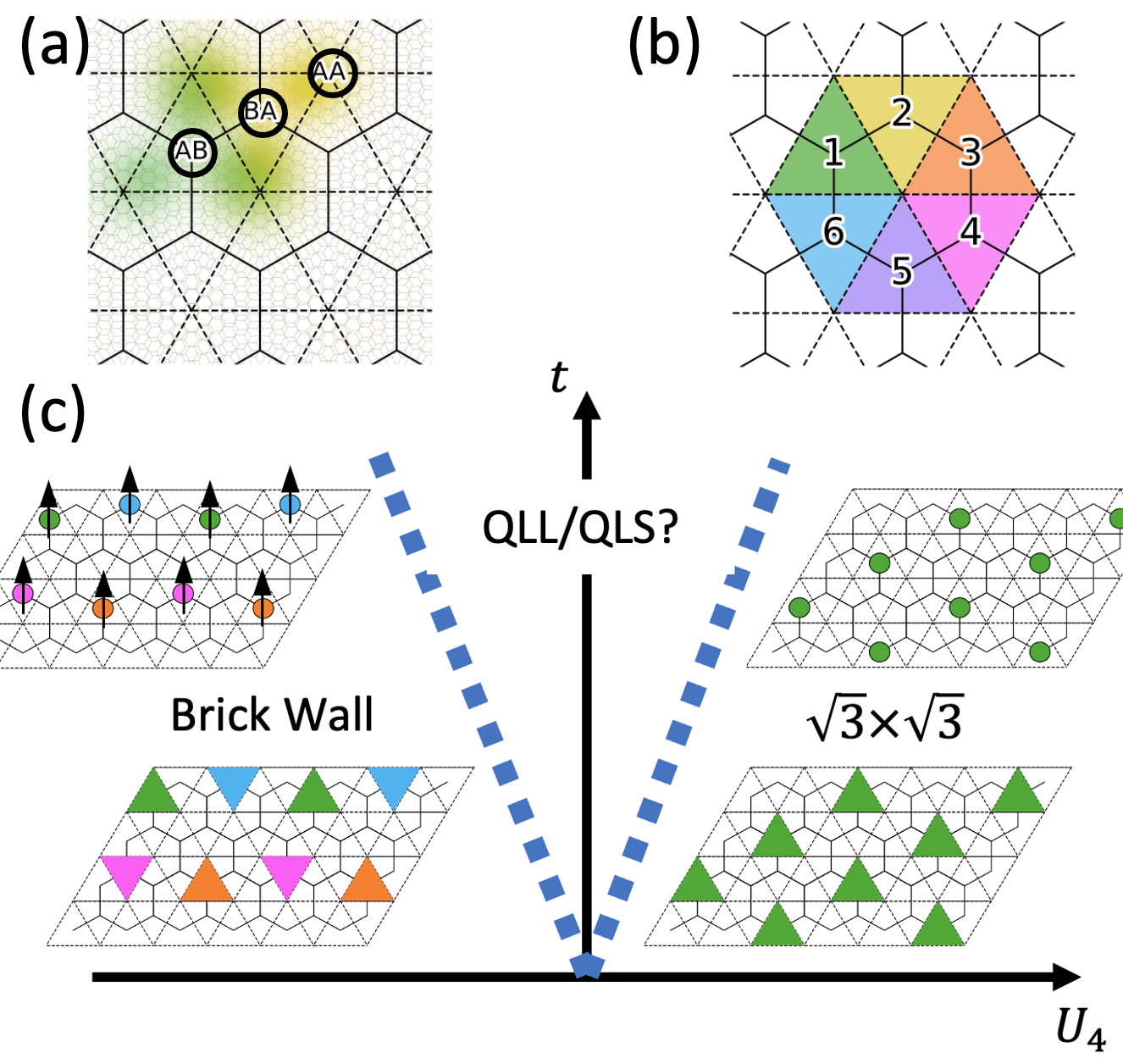}
    \caption{{\bf Wannier states and schematic phase diagram.} (a) Wannier states (WSs) and a typical moir\'{e} pattern.
    The yellow and green blobs schematically represent the shape of Wannier orbitals on the BA and AB sublattices, respectively.
    (b) Schematic representation of the six-phase registry of AB/BA sites. The vertices of the triangle correspond to the three charge lobes.
    (c) Proposed phase diagram for the model in Eq. 3. Black arrows in the brick wall phase represent SU(4) spin-valley ferromagnetism. }
    \label{fig:moire}
\end{figure}

\para 
{\it The Model -- }The topological obstruction forbids symmetric lattice description of the flat bands of magic angle twisted bilayer graphene \cite{Zou2018Phys.Rev.B,Po2018Phys.Rev.X,Koshino2018Phys.Rev.X,Kang2018Phys.Rev.X, ahn2019failure,Song2019Phys.Rev.Lett., liu2019pseudo,Song2021Phys.Rev.B}.
However, the common alignment of twisted bilayer graphene with hexagonal boron nitride (hBN) explicitly breaks the $C_2$ rotational symmetry and justifies construction of Wannier orbitals. Nevertheless, the resulting maximally localized Wannier orbitals are extended beyond their AB/BA site centers \cite{Po2018Phys.Rev.X,Yuan2018Phys.Rev.B} to the three nearest AA sites, with most of the weight equally divided among the three AA sites, forming a ``fidget-spinner'' shape (see \autoref{fig:moire}(a)). Consequently, the dominant interaction term is an on-site repulsive interaction projected to the Wannier orbitals taking a ``cluster-charging" form \cite{motrunich2002exotic,Po2018Phys.Rev.X},
\begin{equation}
    H_U = \frac U2 \sum_r \left(\sum_{i\in \varhexagon_r} n_i\right)^2,
    \label{eq:HU}
\end{equation}
where $\varhexagon_r$ labels the $r$-th hexagonal plaquette and $n_i$ is summed over spin and valley degrees of freedom.

\para We note that the convention in the experimental literature is to view the moir\'e lattice as a triangular lattice with one lattice site per unit cell. On the other hand, the Wannier centers form a honeycomb lattice with two sites per unit cell. 
Hence the conventional filling of $1/3$ electrons or holes for each spin and valley per triangular lattice is equivalent to the filling fraction of $1/6$ per hexagonal lattice per spin and valley (see \autoref{fig:moire}(a)). Hereafter, we refer to such filling as $1/3$ per moir\'e unit cell. At such $1/3$ total filling for spin and valley d.o.f.,
the energy can be minimized by having the charge carriers occupying only one of the six possible registries
(see \autoref{fig:moire}(b). Having $1/3$ charges per moir\'e unit cell corresponds to $\sum_{i\in \varhexagon_r} n_i = 1$ per honeycomb plaquette.)  As pointed out in Ref.~\cite{Zhang2022}, the strong coupling limit (i.e., classical) ground state of  \autoref{eq:HU} is extensively degenerate.

\para Two types of perturbations can lift the extensive ground state degeneracy associated with the cluster-charging interaction of \autoref{eq:HU}: further range interactions and quantum fluctuations.
For TBG systems, the Coulomb interaction projected to low energy Wannier orbitals gives rise to various terms \cite{Po2018Phys.Rev.X, Kang2019Phys.Rev.Lett.,Zhang2019twisted}.
We focus on the $4$th nearest neighbor interactions and  consider the 
 density-density interactions and Hund's coupling to obtain (See Supplementary Material A for detail.)
\begin{equation}
    H_4 = (V_4-V_4^{approx}) \sum_{\langle ij \rangle_4} n_{i} n_{j} -  \frac{J_4}{4} \sum_{\langle ij \rangle_4} (S_i^\mu S_j^\mu + n_i n_j), 
    \label{eq:4-th}
\end{equation}
where $n_i = c_{i\alpha}^\dag c_{i\alpha}$ is the density operator summing over the spin and valley d.o.f. and $S^\mu = c_{i,\alpha}^\dag T_{\alpha\beta}^\mu c_{j,\beta}$ is the $SU(4)$ spin operator, $\alpha, \beta$ denote the combined spin-valley d.o.f. with the $SU(4)$ generators $T^\mu \in \{\sigma^\nu, \tau^{\nu'}, \sigma^\nu \otimes \tau^{\nu'}\}$.
Following the notation of Ref.~\cite{Koshino2018Phys.Rev.X}, $V_4$ ($V_4^{approx}$) is the direct Coulomb interaction between 4th nearest neighbor  (``point-charge-approximated") Wannier orbitals.
The point-charge approximation \cite{Koshino2018Phys.Rev.X} views the fidget-spinner-shaped Wannier orbitals as being composed of three point charges at AA sites.
Focusing on \autoref{eq:4-th} is justified by the fact that the difference between the direct Coulomb interaction and the point-charge approximation is short-ranged while all tiling patterns in the ground state manifold of \autoref{eq:HU} have the same electrostatic potential under the point-charge approximation. 
Finally, $J_4 > 0$ is the $SU(4)$ ferromagnetic exchange interaction ~\cite{Kang2019Phys.Rev.Lett.}.
Upon introducing quantum fluctuations via hopping term $H_K = \sum_{\langle ij\rangle,\alpha,\tau} t_{ij,\tau}( c_{i,\alpha,\tau}^\dag c_{j,\alpha,\tau} + h.c.)$ the full Hamiltonian becomes
\begin{equation}
    H =  H_U + H_4+  H_K.
    \label{eq:ext_hubbard}
\end{equation}
The ground states in the strong coupling limit ($t=0$) was established in Ref.~\cite{Zhang2022}.
With finite hopping $t$, quantum order-by-disorder \cite{Henley1989Phys.Rev.Lett.} would select a different quantum ground state, resulting in a qualitative phase diagram we sketch in \autoref{fig:moire}(c).

\para {\it Strong  Coupling Limit and Fractional Excitations  -- }In the strong coupling limit, the characteristic energy scale is
\begin{equation}
    U_4 = V_4 - V_4^{approx} - \frac{J_4}{2}.
\end{equation}
For $U_4<0$, the system will order into 
 a low-symmetry state dubbed the ``brick wall" \cite{Zhang2022} (\autoref{fig:moire}(c)).
The brick wall tiling makes the maximal use of the Hunds coupling to minimize $H_4$, and will thus be an $SU(4)$ spin-valley ferromagnet.
The anisotropic shape of the mesoscale unit results in low symmetry. Translation, mirror and $C_3$ rotation symmetries of the honeycomb lattice are all broken in the brick wall phase. 
From the point of view of the Wannier orbital centers (circles in \autoref{fig:moire}(c)), the brick wall state is closely related to the
stripe ordered phase  proposed in Ref.~\cite{Kang2019Phys.Rev.Lett.} at filling $n=-3$ of TBG since the brick wall occupies every third sites along a stripe. Hence, the brick wall may be favored at $1/3$ filling away from $n=-3$.
On the other hand, for $U_4>0$, the favored state would be the $\sqrt{3}\times\sqrt{3}$ ordered state, with uniform AB/BA registry.
In this case, from \autoref{eq:4-th}, configurations with different spin-valley orientations are degenerate within the model.
While the two states break translational symmetry in terms of the orbital centers (see filled circles in \autoref{fig:moire}(c)), we anticipate the observable effects of the translational symmetry breaking to be weak due to the spread of the Wannier orbitals. This contrasts the proposed $\sqrt{3}\times\sqrt{3}$ state against the unit-cell tripled charge density wave states proposed in momentum space based numerical approaches~\cite{Wilhelm2021Phys.Rev.B,Zhang2022Phys.Rev.Lett.}.

\para 
A natural consequence of the incompressible tiling in the strong coupling limit at $\pm 1/3$ filling is the possibility of fractionally charged holes.
Intuitively, this can be anticipated by noting
that the 1/3 of electron charge is concentrated at the vertices of the dual triangular lattice for any of the incompressible states~\cite{Koshino2018Phys.Rev.X}. 
The configuration that binds a $1/3$ charge and the energy cost of such an excitation depends on the classical ground state.
However, as we show below, their movements are restricted much like fractons and lineons \cite{Vijay2015Phys.Rev.B,Nandkishore2019Annu.Rev.Condens.MatterPhys.,Pretko2020Int.J.Mod.Phys.A}.

\begin{figure}
    \includegraphics[width=.45\textwidth]{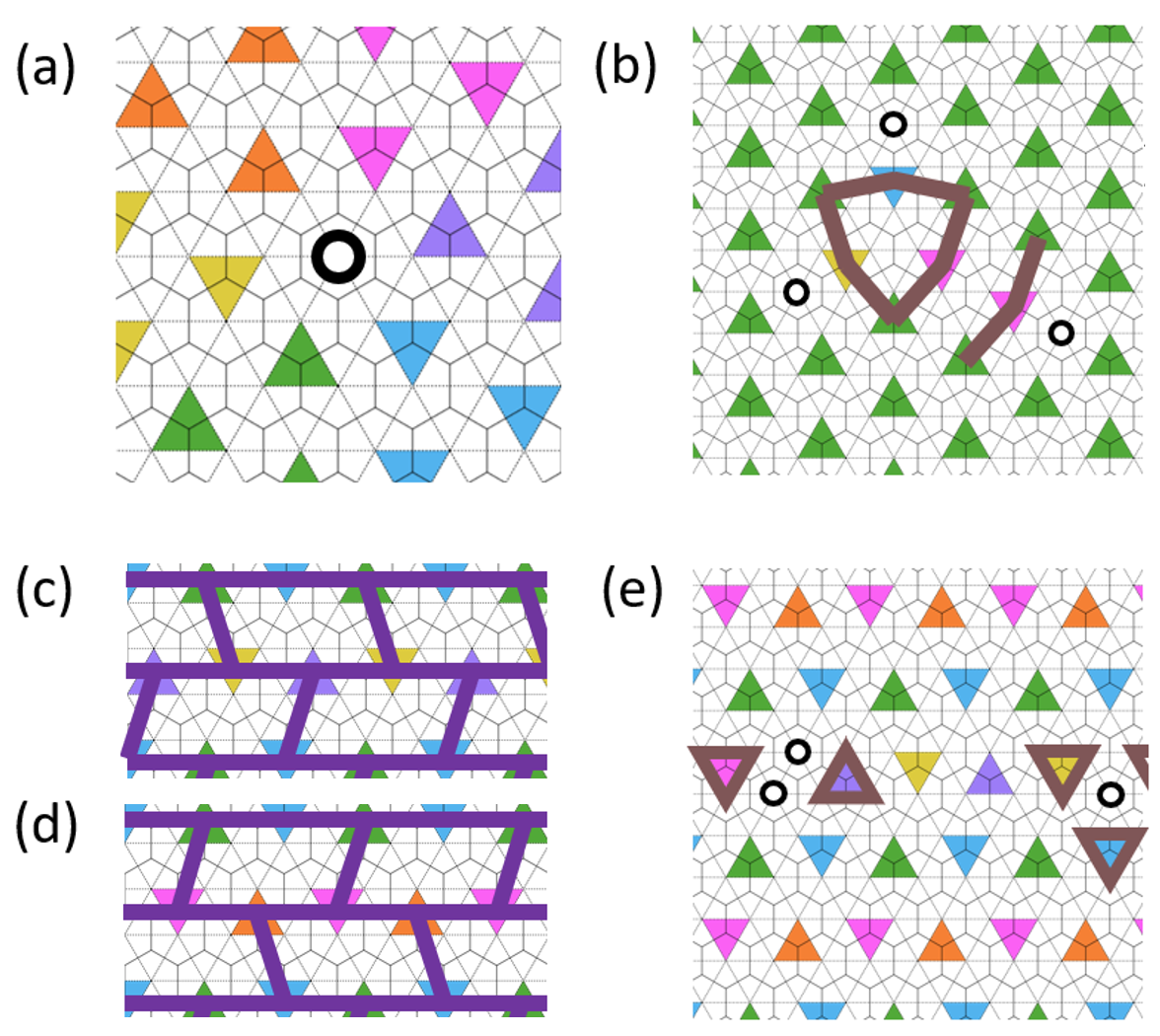}
    \caption{{\bf Fractionally charged excitations.} a) A single $1/3$-charged vacancy (open circle), surrounded by all six registry domains. b) Three solitons with 1/3 charge and flip-line tails. The brown lines indicate the energy cost $U_4$ associated with the domain walls. c-d) Two degenerate brick wall states.  e) Solitons in the brick wall phase can move along a 1D line with constant energy cost associated with brown triangles. }
    \label{fig:domain}
\end{figure}

\para The $\sqrt{3}\times\sqrt{3}$ phase has two types of charge 1/3 fractional excitations with restricted mobility: vortices (\autoref{fig:domain}(a)) and solitons (\autoref{fig:domain}(b)).
As it was previously noted~\cite{Balents2005Phys.Rev.B}, a vortex of phase registry in a charge ordered state usually carries fractional charge.  
An unusual property of our vortices is their restricted mobility: the cluster charging energy $U$ makes the vortices practically immobile, similar to fractons \cite{Vijay2015Phys.Rev.B, vijay2016fracton}. However, due to the extensive enegry cost proportional to $U_4$ associated with the domain walls, the observation of these vortices would require finite temperature.  
We define a ``soliton'' of the $\sqrt{3}\times\sqrt{3}$ phase to be the $1/3$ charged excitation bound to the end of a line of flipped trimers. In the limit of vanishingly small $U_4$, a single hole can fractionalize into three solitons which can only move along one dimension associated with the flip line.
The soliton dynamics are as if a domain wall state of the Su-Schrieffer-Heeger model\cite{Su1979Phys.Rev.Lett.} were embedded in a two-dimensional space.
Hence the soliton behaves like a lineon~\cite{Vijay2015Phys.Rev.B,vijay2016fracton}.
However, the solitons in the $\sqrt{3}\times\sqrt{3}$ state are confined.
The balance between the flip-line energy cost
($2 U_4$ per flip) and the Coulomb interaction between the $1/3$ charges determines the size of the bound state.
From the estimation of $U_4$ in Ref.\cite{koshino2018maximally}, we have $L \sim 1.13 a_M$ (see Supplementary Material B for detailed discussion).

\para Solitons in the brick wall phase are more intriguing because they are deconfined. First we note that as shown in \autoref{fig:domain}(c,d), the brick wall phase has sub-extensive ground state degeneracy since each line of ``bricks" can choose between two degenerate choices of alternating registries that give different slants to the brick tiling pattern.
Hence the ground state degeneracy is $3 \times 2^{L}$ where $L$ is the linear dimension, and the configurational entropy is $ L\log2$ \footnote{The sub-extensive degeneracy is not topologically protected and can be split by further range interactions (see SM F).}.
A defect associated with a domain boundary within a row can also be viewed as a ``soliton" carrying $1/3$ charge (\autoref{fig:domain}(e)) or $2/3$ charge (see Supplementary Material B).
Similar to the $\sqrt{3} \times \sqrt{3}$ phase, the solitons in the brick-wall phase also have restricted mobility, and can only move along the one dimension of the brick wall rows, which are 2D analogs to the ``lineon" excitations in the 3D X-cube model~\cite{Vijay2015Phys.Rev.B, vijay2016fracton}. Furthermore, the solitons in the brick wall phase are deconfined excitations since they cost a finite energy irrespective of the separation between the solitons (see \autoref{fig:domain}(e) and a more detailed illustration in Supplemental Material B).

%{\color{purple} 
The restricted mobility of the solitons seems to happen by chance at first glance. However, as we will show later, these properties are robust against small quantum fluctuations and are closely related to emergent symmetries at low energy. 
%}

\begin{figure}
    \includegraphics[width=.48\textwidth]{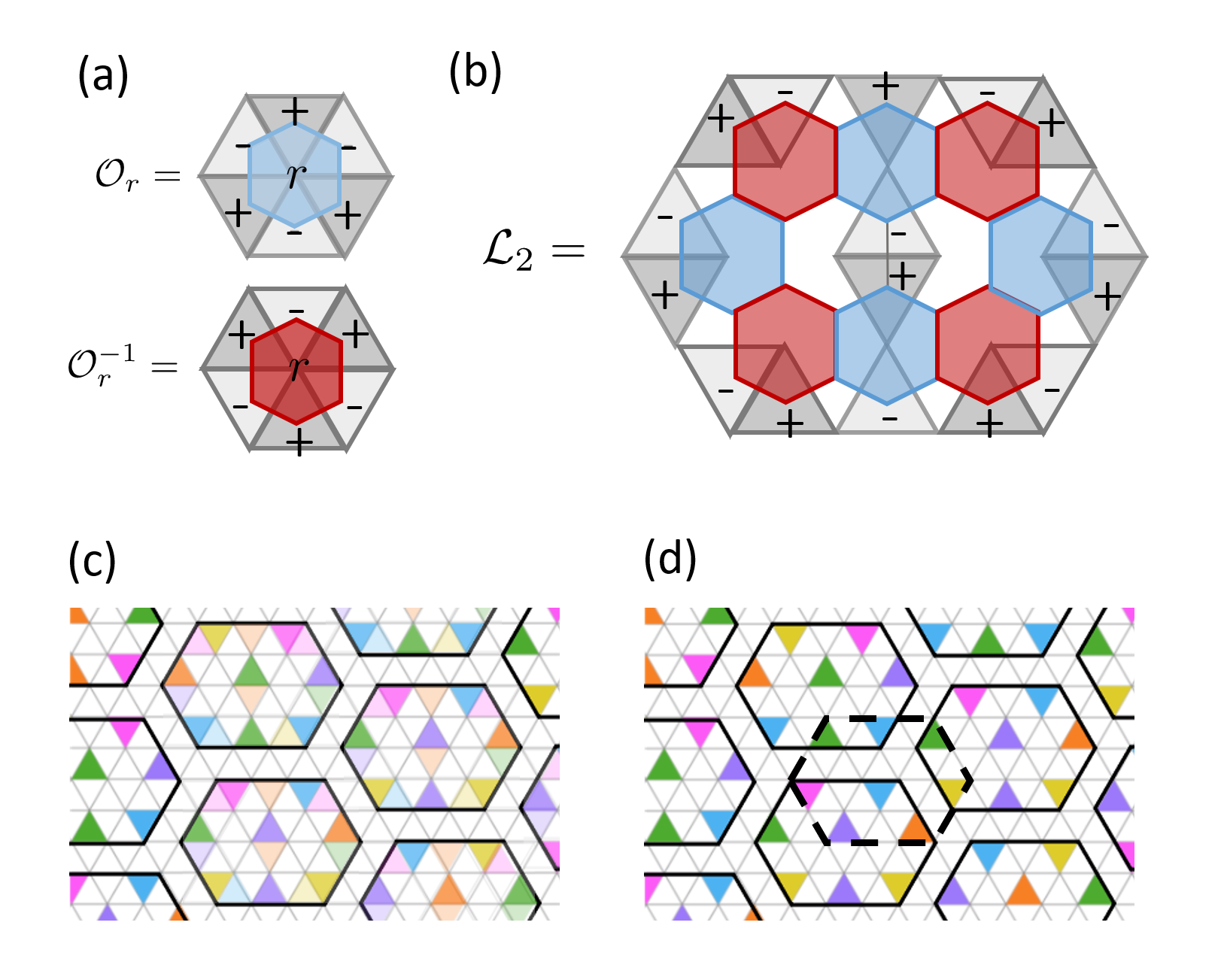}
    \caption{{\bf Operators in effective Hamiltonian \autoref{eq:H_pert} and QLS.}
    a) The lowest order term consists of a plaquette operator $\mathcal{O}_r$. The annihilation (creation) operator acts on the sites marked by -(+) symbols. 
    b) The lowest order nonvanishing operator $L_2$ spans 10 plaquettes.
    c) The plaquette-like candidate state for the QLS, where the resonance happens between the darker shaded triangles and lighter shaded ones.
    d) The columnar-like state. The dotted hexagon denotes an extra flippable pattern when the surrounding three states are all aligned.}
    \label{fig:qls}
\end{figure}

\para {\it Quantum Fluctuations -- }
We now turn to the vertical axis of the phase diagram \autoref{fig:moire}(c) and explore the effects of quantum fluctuations in the limit of $U_4 \ll t$.
We ask how the hopping $t$ in $H_K$ would lift the extensive degeneracy of the $H_U$ ground state manifold through quantum ``order from disorder''~\cite{Henley1989Phys.Rev.Lett.}.
To start answering this question, we look for an operator that can locally connect two different states in the classical ground state manifold.
Such operator should commute with $H_U$, i.e., keep the cluster charge fixed.
Moreover, the operator should act non-trivially in the ground state manifold of $H_U$ at filling $n=\pm1/3$, without annihilating the states in the manifold. 
Since the ground state manifold of $H_U$ at filling $n=\pm1/3$ consists of states with exactly one site of the hexagonal cluster occupied, connecting such states requires coordinated multi-site hopping.
We now show that the smallest such operator consists of an eight-site hopping arranged in a {\it lemniscate}, or sideways figure-eight shape (see \autoref{fig:qls}(b)).

\para
The lemniscate operator $\mathcal{L}_2$ is constructed from the intra-hexagon hopping operator  $\mathcal{O}_{\varhexagon} =  c_{i_1}^\dag c_{i_2} c_{i_3}^\dag c_{i_4} c_{i_5}^\dag c_{i_6}$, where $i_{1,...,6}$ label the sites belonging to a hexagonal plaquette, organized in a clockwise order (\autoref{fig:qls}(a)).
Clearly, $[\mathcal{O}_{\varhexagon}, H_U]=0$, but 
$\mathcal{O}_{\varhexagon} |\psi_0\rangle = 0$ for any $|\psi_0\rangle$ in the ground state manifold of $H_U$ at filling $n = \pm 1/3$ since $\mathcal{O}_{\varhexagon}$ annihilates three fermions on a hexagon, but the cluster-charging constraint requires exactly one fermion on each hexagon.
However, a larger structure involving multiple hexagons built from alternating $\mathcal{O}_{\varhexagon}$ and $\mathcal{O}_{\varhexagon}^{-1}$ operators would still commute with $H_U$ and can be made to be compatible with the cluster-charging constraint.
The lemniscate operator $\mathcal{L}_2$ illustrated in \autoref{fig:qls}(b) is the smallest operator (see Supplemental Material C) that can semi-locally resonate between two different states in the classical ground state manifold.
There are three orientations of lemniscate operators, related by $C_3$ rotation.
For each orientation, the lemniscate operator connects two local tiling configurations, which we designate as the ``flippable" manifold of that operator.
Therefore, the low energy effective Hamiltonian can be written as
\begin{equation}
    \begin{split}
        \Heff &= - \tilde{t} \sum_{i,\alpha} (\mathcal{L}_{2,i,\alpha} + \mathcal{L}_{2,i,\alpha}^\dag) + H_4
    \end{split}
    \label{eq:H_pert}
\end{equation}
where $\tilde{t} \sim t^8/U^7$,  $\alpha\in\{1,2,3\}$ label the three different orientations and $i$ labels the position of the operator $\mathcal{L}_2$ (\autoref{fig:qls}(b)). 

\para 
The effective Hamiltonian $\Heff$ is highly frustrated since different $\mathcal{L}_{2,i,\alpha}$'s do not commute.
Nevertheless, analogies to the quantum dimer models~\cite{moessner2001phase} offer valuable insights. Specifically, 
as in quantum dimer models,
the quantum fluctuations associated with the lemniscate operators would select a novel quantum liquid state or a solid state as a function of $U_4/\tilde{t}$; we refer to these states as 
``quantum lemniscate liquid/solid" (QLL/QLS)(\autoref{fig:moire}(c)). 
%{\color{purple} 
The quantum fluctuation through the lemniscate operators will avoid any mobility restriction for doped charges in both phases.
%}

Among possible QLS states are a plaquette-like state that has resonance within supercells (\autoref{fig:qls}(c)) and a columnar-like state with fixed configurations within supercells that repeats for $U_4 < 0$ (\autoref{fig:qls}(d)) or alternates for $U_4>0$.
Both the plaquette-like and columnar-like QLS states break $C_3$ rotational symmetry in addition to the lattice translation symmetry. The supercells act as an emergent local degrees of freedom analogous to the emergent orbitals in the so-called cluster Mott insulators on the kagome lattice \cite{Chen2018Phys.Rev.B}.
However, while such emergent orbitals are pinned to the lattice, our supercells form an emergent superstructure in the QLS.
Doping away from $1/3$ filling, holes added to the QLS phases can also fractionalize into $1/3$-charged excitations.
%{\color{purple} 
However, they are energetically confined as in the $\sqrt{3}\times\sqrt{3}$ phase (see SM C).
%}

\para %{\color{purple}
While mapping out the conditions for the QLL ground state of the Hamiltonian Eq.\ref{eq:H_pert} would require numerical or quantum simulation of the model, some properties of a QLL state can be anticipated on general grounds. 
One mechanism that would favor a QLL over a QLS state is through resonances unconstrained to a rigid cell (e.g. the dotted hexagon in \autoref{fig:qls}(d)). As we describe using a minimal effective model in SM D, such resonance will promote a QLL state that breaks the $C_3$ rotational symmetry. 
A gapped and translationally invariant QLL state must host deconfined charge 1/3 anyonic excitations based on LSM(Lieb-Schultz-Mattis)-type constraints~\cite{Cheng2016Phys.Rev.X}. Such excitations can be viewed as the 1/3-charged lineons becoming fully mobile due to lemniscate resonances. 
Alternatively, a QLL analogous to the valence bond liquid state at  the the so-called 
Rocksar-Kivelson (RK) point ~\cite{Rokhsar1988Phys.Rev.Lett.,moessner2001resonating} would be 
 an equal weight superposition of all the possible tiling configurations. Such a QLL state will respect $C_3$ rotation symmetry.
%}

%{\color{purple}
\para {\it Emergent 1-form symmetry} -- 
We turn to the theoretical implications of the fractonic restricted mobility of our fractionally charged defects. It is believed that fracton phases do not exist in fully gapped systems in two spatial dimensions (2D) \cite{haah2011local}, without symmetry protection. However subsystem symmetry \cite{qi2021fracton,shen2022fracton,rayhaun2021higher} or multipolar symmetry \cite{Pretko17, Pretko17generalized,Pretko18} can result in fractonic excitations with restricted mobility in 2D\cite{mcgreevy2023generalized}. 
Curiously, we find the mobility restriction of fractional charge in our model is tied to a new emergent 1-form symmetry; a new mechanism for fractonic excitations in 2D.

The notion of $p$-form symmetry, symmetry operator acting on codimension-($p+1$) submanifold of the spacetime, has garnered interest in the community as a framework that unifies Landau symmetry breaking paradigm with topologically ordered phases\cite{mcgreevy2023generalized}.  
As we prove in SM.F, the cluster charging constraint implies an emergent $U(1)\times U(1)$ 1-form symmetry at low energy. The string operator charged under the symmetry moves the fractionally charged excitation from one end to the other. 
This 1-form symmetry unifies the distinct descriptions of fractional charge motion as follows.  In the 
 brick-wall and $\sqrt{3}\times \sqrt{3}$ phases, the string operators are rigid, 
 resulting in restricted mobility of the lineons.  Contrastingly, in the QLL phase the string operator is allowed to fluctuate, resulting in unrestricted motion of the fractional charge.

%}

\para {\it Experimental Implications --}
Our rich phase diagram with exotic states in experimentally accessible platform opens door for detection and control of novel states.
%{\color{purple} 
 The restricted mobility of lineons in the brick-wall phase gives rise to emergent Luttinger liquid behavior at small hole doping away from filling of $1/3$. The lineon motion can be modeled using three flavors of solitons.
In contrast to the well-studied commensurate-incommensurate transition near $1/3$ filling in one-dimension predicted to exihibit the Luttinger parameter $K=1/9$ \cite{giamarchi2003quantum}, we predict the Luttinger parameter the emergent lineon Luttinger liquid to be $K=1/3$  (see SM. E).
 The prediction can be verified through Luttinger liquid scaling of conductance and a violation of Wiedmann-Franz law and divergent Lorentz number at low temperature (see SM. E). 
 
More broadly, our formalism can be generalized and applied to other third fillings by accommodating more electrons per honeycomb plaquette (see SM. G). Furthermore, since the geometry of the extended orbital does not require the fine-tuning of the magic angle, we anticipate the fractional incompressible states at $n\pm 1/3$ to be robustly present even at larger twist angles
\cite{Jeanie_unpub}. 

For fillings larger than $1$, some sites will have double occupation, resulting in a competition between spin-singlet, valley-polarized and spin-triplet, valley-anti-aligned states. Switching between competing states will manifest through non-monotonic magnetotransport under an in-plane field. 

Finally, mirror-symmetric twisted trilayer graphene at 1/3 filling can host a fractional correlated insulating state presented in this letter with an additional Dirac cone at charge neutrality (see SM. H). Interestingly, recent experiments on 
 twisted tri-layer graphene reported observation of zero Chern number incompressible states\cite{shen2023dirac}.

%}

\begin{acknowledgements}
{\bf Acknowledgements} 
We thank L. Balents, F. Burnell, O. Vafak, S. Vijay, C.N. Lau, M.W. Bockrath, Z. Bi, C.-M. Jian, Y. You, Y.-H. Zhang, T. Senthil, R. Nandkishore and M. Hermele for illuminating discussions and helpful comments. DM was supported by the Gordon and Betty Moore Foundation’s EPiQS Initiative, Grant GBMF10436. KZ was supported by NSF EAGER OSP\#136036 and NSERC.
E-AK acknowledges funding through Simons Fellows in Theoretical Physics award number
920665 and by the Ewha Frontier 10-10 Research Grant. 
Part of this work was performed at the Aspen Center for Physics, which is supported by National Science Foundation grant PHY-160761.
\end{acknowledgements}
\bibliographystyle{apsrev4-2}
\bibliography{Moire-fracton}
\end{document}

\title{Supplementary Information - Fractionalization in Fractional Correlated Insulating States at $n\pm 1/3$ filled twisted bilayer graphene}

\author{Dan Mao}
\author{Kevin Zhang}
\affiliation{Laboratory of Atomic and Solid State Physics, Cornell University, 142 Sciences Drive, Ithaca NY 14853-2501, USA}

\author{Eun-Ah Kim}
\affiliation{Laboratory of Atomic and Solid State Physics, Cornell University, 142 Sciences Drive, Ithaca NY 14853-2501, USA}
\date{October 2021}
\affiliation{Radcliffe Institute for Advanced Study at Harvard, Harvard University, 10 Garden Street, Cambridge MA 02138, USA}
\affiliation{Department of Physics, Harvard University, 17 Oxford Street, Cambridge MA 02138, USA}
\affiliation{Department of Physics, Ewha Womans University}

\maketitle

\section{A. Contributions to 4th nearest neighbor density-density interaction}

We first describe the direct interaction term.
The total electrostatic potential can be written as $V = \sum_{m=4}^\infty \sum_{\langle ij \rangle_m} V_m \hat{n}_i \hat{n}_j$, where $V_m$ is the direct Coulomb interaction between the $m$th nearest neighbor Wannier orbitals.
The sum starts at $m=4$ as the no-touching constraint forbids smaller $m$.
Similar to the Ewald summation \cite{ewald}, we split the electrostatic potential $V$ into a short-range part and a long-range part,
\begin{equation}
\begin{split}
    V &= \sum_{m=4}^\infty \sum_{\langle ij \rangle_m} (V_m -  V_m^{approx} )\hat{n}_i \hat{n}_j  + \sum_{m=4}^\infty \sum_{\langle ij \rangle_m}  V_m^{approx} \hat{n}_i \hat{n}_j\\
    &\equiv V_{short} + V_{long},
\end{split}
\end{equation}
where $V_m^{approx}$ is the Coulomb interaction between the $m$th-nearest neighbor ``point-charge-approximated" Wannier orbitals. The point-charge approximation \cite{Koshino2018Phys.Rev.X} views the fidget-spinner-shaped Wannier orbitals as three point charges at AA sites. As a result, configurations satisfying the cluster-charging condition have the same electrostatic potential, so $V_{long}$ is a constant.
Therefore, we only need to consider $V_{short}$.
Since $V_m -  V_m^{approx}$ decays quickly with increasing $m$ \cite{Koshino2018Phys.Rev.X}, we only keep the leading order contribution, which is $m=4$.

Other than the direct Coulomb interaction, we also take into account the exchange interaction.
From the projection of the Coulomb interaction onto the Wannier basis, the exchange term can be written as
\begin{equation}
\begin{split}
    H_{exchange} = \sum_{i\neq j,\alpha,\beta,\tau,\tau'} J_{ij}^{\tau \tau'} c_{i,\alpha,\tau}^\dag c_{j,\alpha,\tau} c_{j,\beta,\tau'}^\dag c_{i,\beta,\tau'},
    \end{split}
    \label{eq:exchange}
\end{equation}
where $i,j$ represent Wannier centers, $\alpha,\beta \in\{\uparrow,\downarrow\}$ label the spin, $\tau,\tau' \in\{+,-\}$ label the valley and $V_{ij}^{\tau \tau'} = \sum_{r,r'} V(|r-r'|) \psi_{i,\tau}^*(r) \psi_{j,\tau}(r) \psi_{j,\tau'}^*(r') \psi_{i,\tau'}(r')$ with $V(r-r')$ being the Coulomb potential and $\psi_{i,\tau}(r)$ being the Wannier function. We ignore the inter-valley Hund's interaction since it is negligible compared to \autoref{eq:exchange} \cite{Koshino2018Phys.Rev.X}. We also assume approximate $SU(4)$ symmetry of the spin-valley d.o.f., that is, $J_{ij}^{+ +} \approx J_{ij}^{- -} \approx J_{ij}^{ + -} \approx J_{ij}^{- +} \equiv J_{ij}$ \cite{Kang2019Phys.Rev.Lett.}.
Therefore, the leading order exchange interaction with $J_4$ being defined as $J_{ij}$ for 4th nearest neighbors $i, j$ is
\begin{equation}
\begin{split}
&J_4 \sum_{\langle i,j\rangle_4} c_{i,\alpha,\tau}^\dag c_{j,\alpha,\tau} c_{j,\beta,\tau'}^\dag c_{i,\beta,\tau'}\\
=& - \frac{J_4}{4} \sum_{\langle i,j\rangle_4} S_i^\mu S_j^\mu - \frac14 J_4 \sum_{\langle i,j\rangle_4} n_i n_j,
\end{split}
\label{eq:J4}
\end{equation}
where $S_i^\mu = c_{i,\eta}^\dag T_{\eta \eta'}^\mu c_{i,\eta'}$ and the repeated indices are summed over. $\eta, \eta' \in \{1,2,3,4\}$ denote the combined spin-valley indices and $T^\mu$ is the $SU(4)$ generator with $\mu \in\{1,...,15\}$, which we choose to be $\{\sigma^\nu, \tau^{\nu'}, \sigma^\nu \otimes \tau^{\nu'}\}$, where $\sigma^\nu$, $\tau^{\nu'}$ are Pauli matrices. The last equality in \autoref{eq:J4} follows from the completeness relation of $SU(2)$, that is $\sigma_{\eta_1 \eta_2}^{\nu} \sigma_{\eta_3 \eta_4}^{\nu} = 2 \delta_{\eta_1 \eta_4} \delta_{\eta_2 \eta_3} -  \delta_{\eta_1 \eta_2} \delta_{\eta_3 \eta_4}$.

\section{B. Fractional excitations in $\sqrt{3} \times \sqrt{3}$ and brick wall phases}

\begin{figure}
    \centering
    \includegraphics[width=0.48\textwidth]{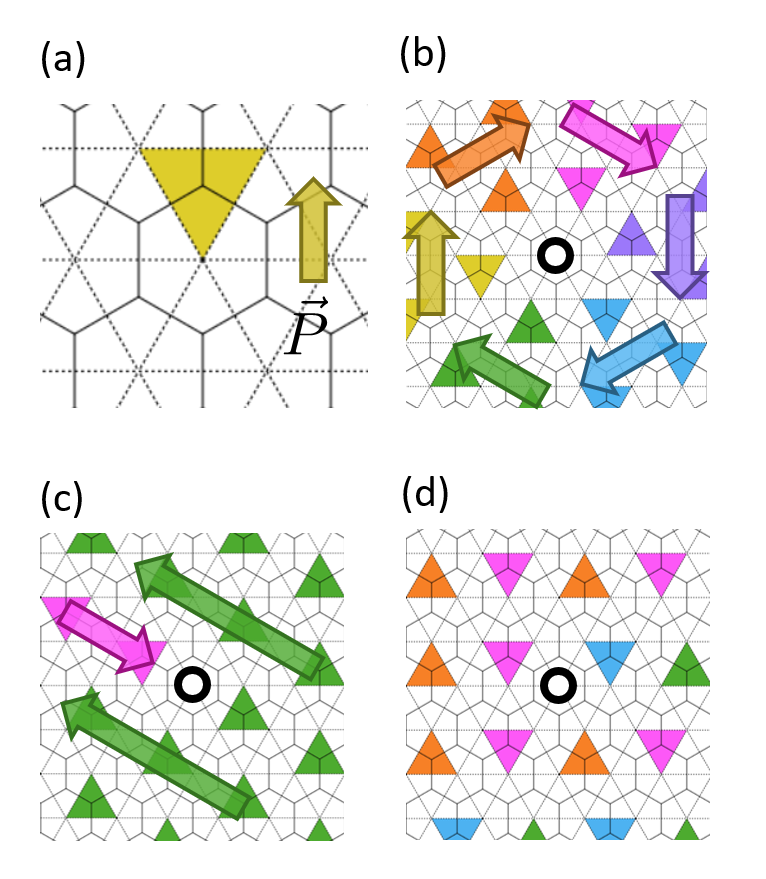}
    \caption{a) A single trimer and its associated polarization direction.
    b) A soliton excitation in the $\sqrt{3} \times \sqrt{3}$ phase consists of a trail with differing registry from a uniform background.
    c) A soliton in the brick wall phase is surrounded by various differing registry domains.}
    \label{fig:polarization}
\end{figure}

In this section, we describe the restricted mobility of the vortex and soliton in the $\sqrt{3} \times \sqrt{3}$ and brick wall phases and their confinement properties.
We first define the fractional excitations as follows.
By assigning a polarization direction $\Vec{P}$ to each of the six phase-registries (\autoref{fig:polarization}(a)), we can calculate the local charge of a tiling configuration as the bound charge $q_b = \oint d\vec{l} \wedge \Vec{P}$.
A vortex excitation of the $\sqrt{3}\times\sqrt{3}$ state can thus be seen as the excitation resulting from a vortex of the polarization direction (\autoref{fig:polarization}(b)), with $1/3$ bound charge.
The solitons of the $\sqrt{3}\times\sqrt{3}$ and brick wall state can be viewed similarly (\autoref{fig:polarization}(b,c)), although the brick wall soliton is considerably more complex in terms of registry polarization.
We now discuss the properties of these excitations individually.

\subsection{Charge $1/3$ vortex in $\sqrt{3}\times\sqrt{3}$}

\begin{figure*}
    \centering
    \includegraphics[width=0.96\textwidth]{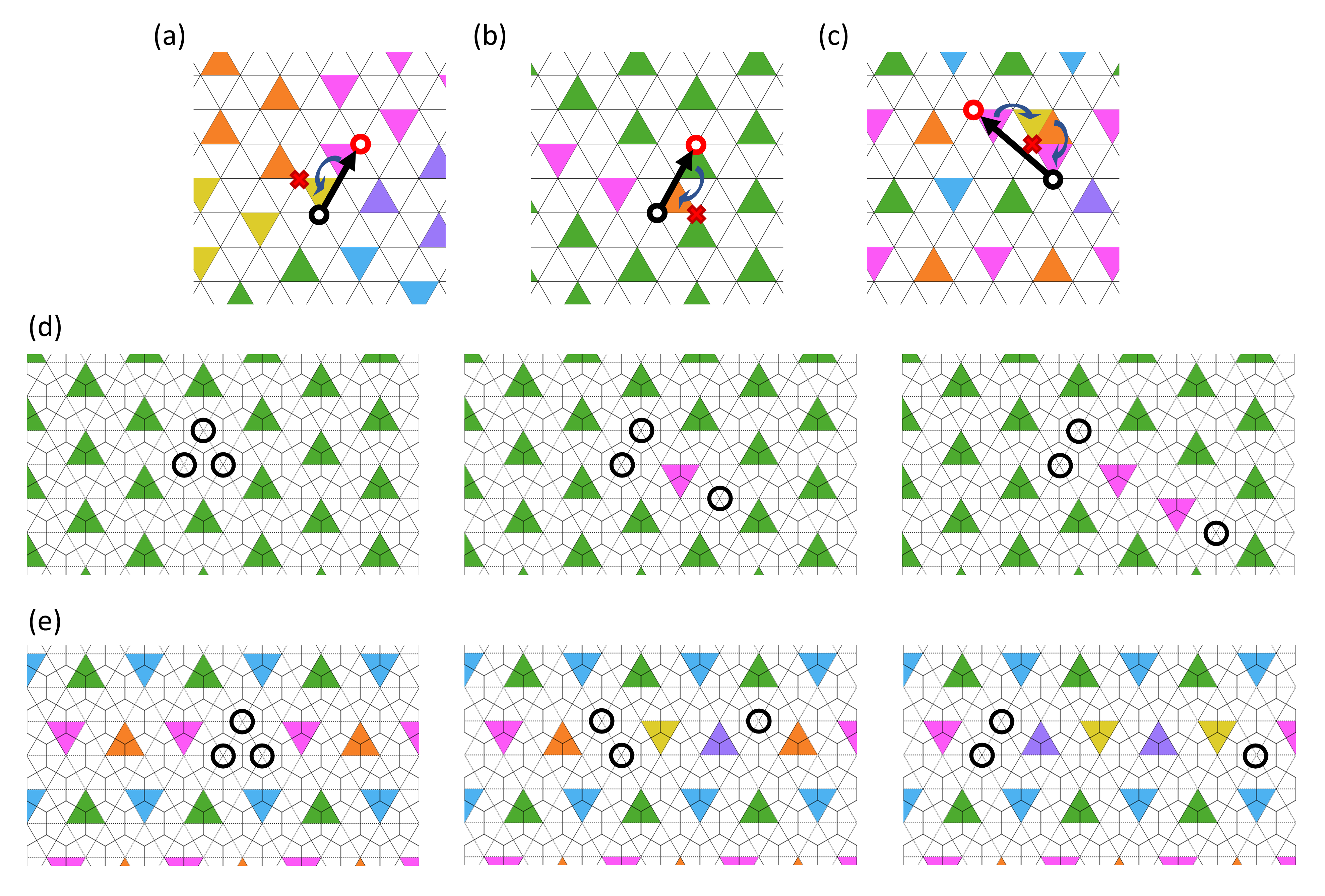}
    \caption{Restricted mobility of excitations. a) Attempting to move a vortex quasiparticle from the black to the red ring by flipping the trimer marked with an arrow would cause a double occupancy shown by the red cross. b) Moving a soliton outside of the allowed direction (upper-left to bottom-right diagonal) in the $\sqrt{3} \times \sqrt{3}$ state would cause a double occupancy. c) Moving a soliton outside of the allowed direction (horizontal) in the brick wall phase requires two flips, but would still cause a double occupancy.
    d) Soliton movement in the $\sqrt{3} \times \sqrt{3}$ phase.
    e) Soliton movement in the brick wall phase. The initial hole fractionalizes into a left-moving soliton with a charge $2/3$ and a right-moving soliton with a charge $1/3$.}
    \label{fig:restriction}
\end{figure*}

\autoref{fig:restriction}(a) shows an immobile vortex quasiparticle, marked by the black empty circle.
The vortex is surrounded by all six phases of the AB/BA site registry and has a charge $1/3$.
Attempting to move this quasiparticle to the location marked by the red empty circle, by flipping the trimer indicated by the arrow, would result in a double occupancy shown by the red cross.
Any different attempt to move the vortex would also result in a similar double occupancy.

\subsection{Confined charge $1/3$ soliton in $\sqrt{3}\times\sqrt{3}$}

A soliton here has a charge $1/3$ and can move along a one-dimensional line.
It is a boundary between the two phases of the AB/BA site registry.
In \autoref{fig:restriction}(b), the mobility direction is shown by the line of flipped pink triangles, and the soliton acts as a boundary between pink and green phases.
Attempting to move the soliton outside of this predefined direction (for example, perpendicular to its mobility direction, to the red empty circle) would, as above, result in double occupancy (red cross).
A hole with charge $1$ would generate three solitons, each of which can move independently.
\autoref{fig:restriction}(d) shows a single hole and the individual steps taken in moving one of the three 1/3 charges, flipping one triangle at a time.
The three solitons move along different axes related by $C_3$ symmetry, where shown movement is along the upper-left to bottom-right diagonal.

We now discuss the confinement of the solitons.
Firstly, their movement leaves behind a trail in the background electronic configuration.
This can be seen in \autoref{fig:confinement}(a) as trails of blue, magenta, or yellow flipped triangles.
Each trimer flipped by the movement of a soliton is associated with a short-range energy cost of $2 U_4$ per unit length of $\sqrt{3} a_M$(crimson bonds), and since the number of flips increases with the traveled distance of the solitons, they are confined in this phase.
There is also long-range Coulomb repulsion of the charges.
Hence, the total energy cost for a three-soliton configuration with linear dimension $l$ is given by $E(l) = 3 \left(2 U_4 \frac{l}{\sqrt{3} a_m} \right) + 3\frac{e^2}{9\epsilon a_m (1 + \sqrt{3} l)}$, where $\epsilon$ is the dielectric constant and $a_M$ is the moir\'e lattice constant. By minimizing $E(l)$, we obtain the confinement length $L$.
For the solitons to have finite confinement length, i.e. $L>0$, $U_4 < \frac{e^2}{6 \epsilon a_M}$ must hold.
The confinement length is $L = \frac{1}{\sqrt{3}}\left[\sqrt{\frac{a_M e^2}{6 \epsilon U_4}} - a_M \right] $. 

\subsection{Deconfined charge $1/3$ and $2/3$ solitons in brick-wall}

\begin{figure*}
    \includegraphics[width=.96\textwidth]{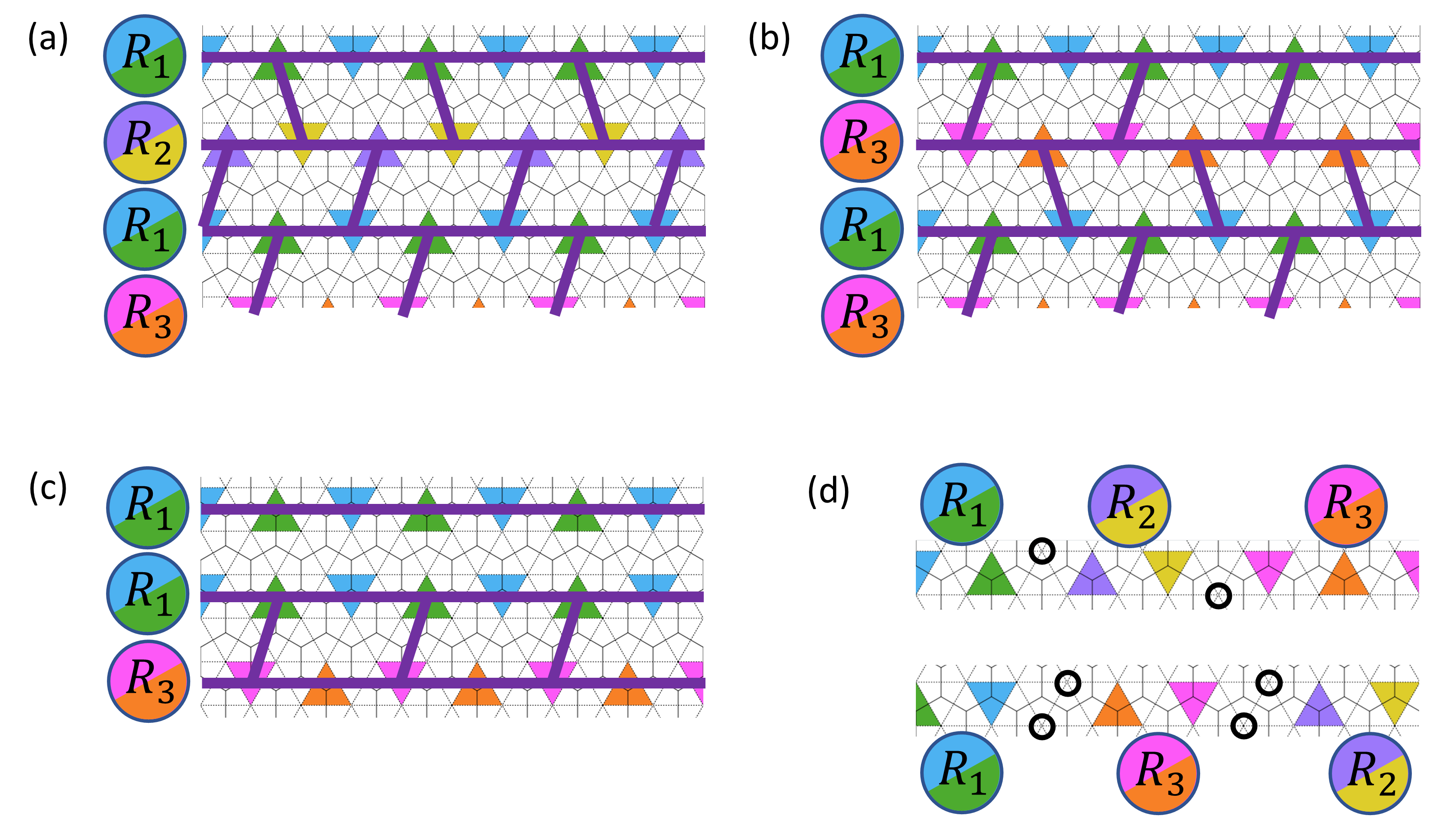}
    \caption{Three domains of trimer rows in the brick-wall phase.
    a,b) Each row must have a differing phase from the rows above and below.
    The residual freedom manifests as sub-extensive entropy, which can be seen through the freedom in choosing ``slant directions" for each row of bricks.
    c) Two adjacent rows with the same phase would not be able to form $U_4$ bonds, and is thus suboptimal in energy.
    d) $1/3$ charge solitons are domain walls between two of the three phases. Likewise, $2/3$ charge solitons are also domain walls, but with the opposite direction of cycling between the three phases.}
    \label{fig:brickphases}
\end{figure*}

The brick-wall phase also hosts solitons (\autoref{fig:restriction}(e)) but with different mobility and confinement properties.
Due to the reduced symmetry of the brick-wall phase, the nature of the solitons is significantly more complex than the $\sqrt{3} \times \sqrt{3}$ case.
We first define the soliton by analyzing the constraints on the ground-manifold of the brick-wall state.
Without loss of generality, we consider rows of bricks that are horizontally aligned (\autoref{fig:restriction}(f)).
In this state, the trimers also form rows of alternating AB/BA (up/down) centers.
There are three possible ``phases" for the rows: $R_1$ (cyan/green), $R_2$ (purple/yellow), and $R_3$ (magenta/orange).
In order to maximize the number of $U_4$ bonds per trimer, each row must have a phase differing from those of the rows above and below.
Two adjacent rows with the same phase would not form any $U_4$ bonds (\autoref{fig:brickphases}(c)), and thus be suboptimal in energy.
This constraint is the origin of the subextensive entropy of the brick wall state.
If one ``fills up" the lattice sequentially from top to bottom with rows of trimers, there will always be two allowed phases for each row (\autoref{fig:brickphases}(a,b)).

A soliton, then, can be viewed as a boundary between two phases within a row (\autoref{fig:brickphases}(d)).
A $1/3$ charge soliton is a boundary between $R_1$/$R_2$, or $R_2$/$R_3$, or $R_3$/$R_1$ (viewed from left to right).
Two $1/3$ charge solitons can also join together to form a $2/3$ charge soliton, which is the boundary between $R_1$/$R_3$, or $R_2$/$R_1$, or $R_3$/$R_2$ (again from left to right).
The solitons can move horizontally by flipping trimers.
However, like earlier, any movement of a soliton outside of the mobility direction would result in double occupancy (\autoref{fig:restriction}(c)).
Here, the movement would require two nearest-neighbor flips (denoted by the two arrows).
\autoref{fig:restriction}(e) shows a single hole, fractionalized into a $2/3$ soliton moving to the left, and a $1/3$ soliton moving to the right.
Unlike the $\sqrt{3} \times \sqrt{3}$ phase, all solitons move along the same axis.

Remarkably, the solitons shown in \autoref{fig:restriction}(e) are deconfined.
Earlier, confinement arose from the solitons' movement leaving behind trails of flipped trimers that cost energy via $U_4$.
Analogously, here we ask the question of whether the solitons' movement costs energy scaling with distance.
As established earlier, solitons separate domains of $R_1$, $R_2$, or $R_3$.
Fractionalization of a single hole would result in at least two different phases existing within the same row.
Therefore, if the rows above and below have differing phases, the solitons would leave behind a trail of flipped trimers that would be in the same phase as one of the rows above or below and cost an average energy of $|U_4|$ per two flips (unit length of $3a_M/2$).
The total energy cost here for a three-soliton configuration where the central $1/3$ charge remains stationary is approximately $E(l) = 2 |U_4| \frac{l}{3a_M} + \frac{5}{2}\frac{e^2}{9\epsilon a_M l}$.
The confinement length here, obtained by minimizing $E$, would be $L = \sqrt{\frac{5a_m e^2}{12|U_4| \epsilon}}$.

\begin{figure}
    \includegraphics[width=.48\textwidth]{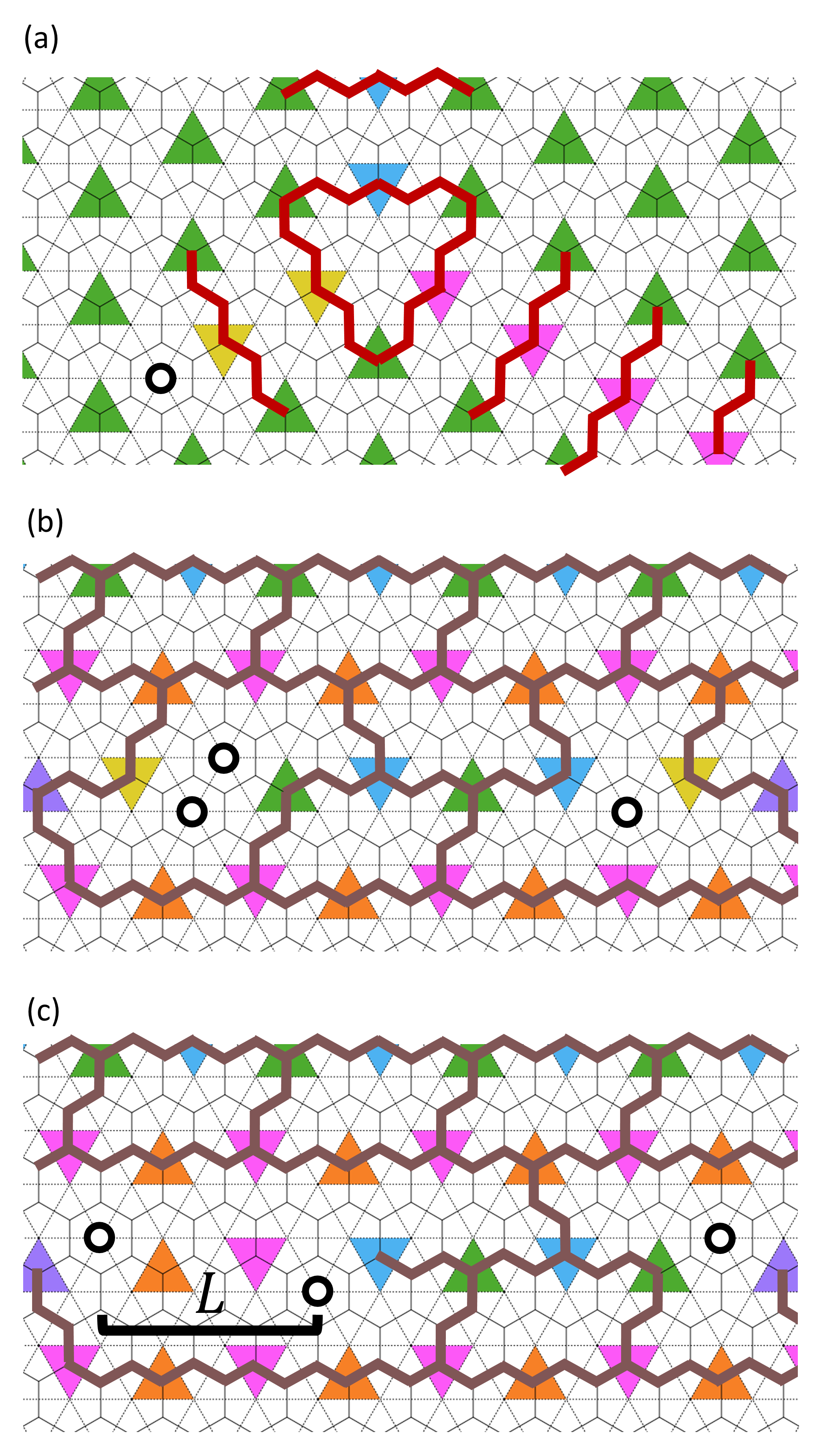}
    \caption{Confinement cost of solitons in the $t=0$ phases. a) In the $\sqrt{3} \times \sqrt{3}$ phase, the trail of flipped trimers caused by the movement of a soliton costs energy of $2 U_4$ per unit length (red lines show $U_4$ bonds which cost energy).
    b) In the brick wall phase, the trail of flipped trimers is able to form $U_4$ bonds with the surrounding $R_3$ phase above and below (brown lines show $U_4$ bonds which lower the energy).
    The two solitons enclose a new $R_1$ domain in the middle of the $R_2$ phase, both of which satisfy the $U_4$ constraint.
    c) The $2/3$ charge soliton could separate to form a bound state with finite size $L$ due to Coulomb repulsion, but the bound state remains deconfined.}
    \label{fig:confinement}
\end{figure}

On the other hand, if the rows above and below the hole have the same phase, there can be two different domains that satisfy the $U_4$ constraint.
This suggests that the hole could fractionalize into a pair of $1/3$ charge and $2/3$ charge solitons that are boundaries between the two allowed phases.
\autoref{fig:confinement}(b) shows one such case of a hole in a $R_1$ row, where the rows above and below are $R_3$.
The $2/3$ charge and $1/3$ charge solitons form left and right boundaries respectively for the central region of $R_2$.
In this case, the $2/3$ soliton must be on the left side; if it were on the right, the central domain would be $R_3$ and therefore not allowed.
In general, the direction of the $2/3$ soliton's move depends on the surrounding domains.
Due to Coulomb repulsion, the individual quasiparticles that make up the $2/3$ soliton would separate somewhat, leading to a bound state of two $1/3$ solitons (\autoref{fig:confinement}(c)) with length scale $L=\sqrt{\frac{a_M e^2}{6 |U_4| \epsilon}}$ obtained by minimizing the energy $E(l) = |U_4| \frac{l}{3a_m/2} + \frac{e^2}{9\epsilon a_M l}$.
Still, the $2/3$ charge bound state as a whole is deconfined.
Overall, since the energy cost (broken $U_4$ bonds in \autoref{fig:confinement}(b,c)) is localized to a finite area around all of the solitons, they are deconfined.

\section{C. Effective Hamiltonian of lemniscates}

\begin{figure}
    \includegraphics[width=.48\textwidth]{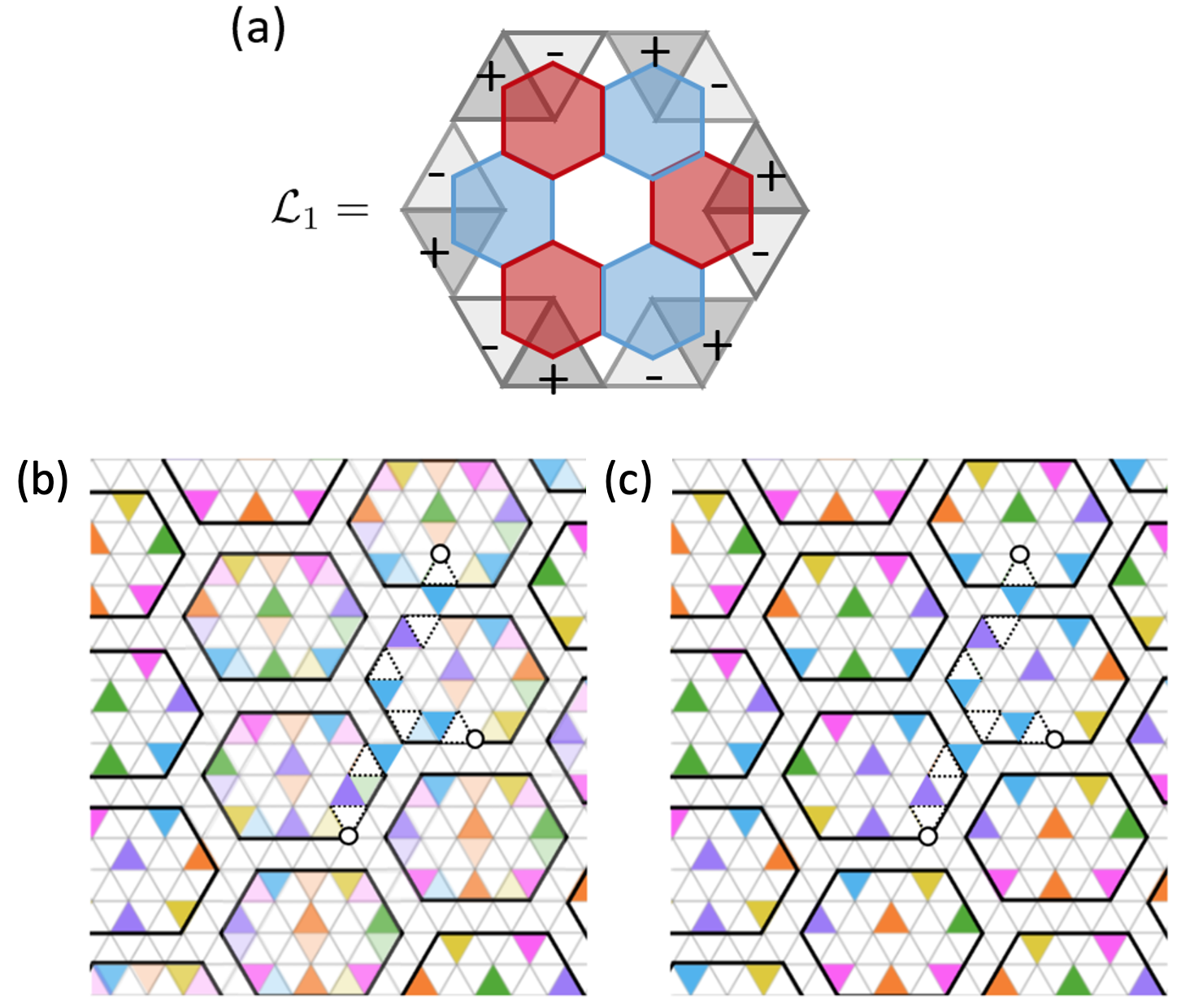}
    \caption{a) The lowest order operator $L_1$, which contains only one creation and one annihilation operator on the supported hexagonal plaquettes, spans seven plaquettes, with nontrivial annihilation and creation operators acting on the outermost 12 sites. The middle plaquette must violate the cluster-charging constraint (containing a vortex) given that the surrounding plaquettes satisfy the constraints.
    b,c) Typical configurations of fractionalized charge excitations for plaquette-like order and columnar-like order, respectively.
    The dotted triangles denote the originally occupied sites and the black circles denote the 1/3-charged excitations.}
    \label{fig:l1}
\end{figure}

In this section, we explain in detail why $\mathcal{L}_2$ is the lowest-order operator that acts non-trivially on the classical ground state manifold of the no-touching model.

As explained in the main text, the necessary condition for a local operator to act non-trivially on the low-energy Hilbert space is to contain at most a pair of creation and annihilation operators per plaquette. The lowest order of such an operator is a single loop of $\mathcal{O}_{\varhexagon}$s and $\mathcal{O}_{\varhexagon}^{-1}$s, which we define as $\mathcal{L}_1$ (\autoref{fig:l1}(a)). However, $\mathcal{L}_1$ would only act non-vanishingly on a configuration with a vortex in the center.
Since vortices are excitations, this operator would annihilate any configuration which satisfies the no-touching condition at $1/3$ filling.
Thus, the next lowest order operator is $\mathcal{L}_2$, which we have considered in the main text.

\section{D. Properties of QLS and competing phases}
\subsection{Fractional charged excitaion in QLS}
The QLS states can also host fractionalized excitations with charge 1/3(\autoref{fig:l1}(b,c)). Typical configurations of three fractionalized charge excitations are illustrated in \autoref{fig:l1}(b,c) for plaquette-like order and columnar-like order.
As in the $\sqrt{3}\times\sqrt{3}$ ordered phase, movements of these $1/3$-charged excitations necessarily disrupt ``flippable" configurations defining the supercells that are rigidly arranged in the QLS phase, and therefore they are confined excitations.

\subsection{Effective model for competing phases near QLS}
Although QLS is more favored by quantum fluctuation compared to the brick-wall and $\sqrt{3}\times\sqrt{3}$ phase, lemniscate resonance in the QLS can lead to melting of the rigid arrangement of the supercells and give rise to a possible QLL state.

\begin{figure}
    \includegraphics[width=.48\textwidth]{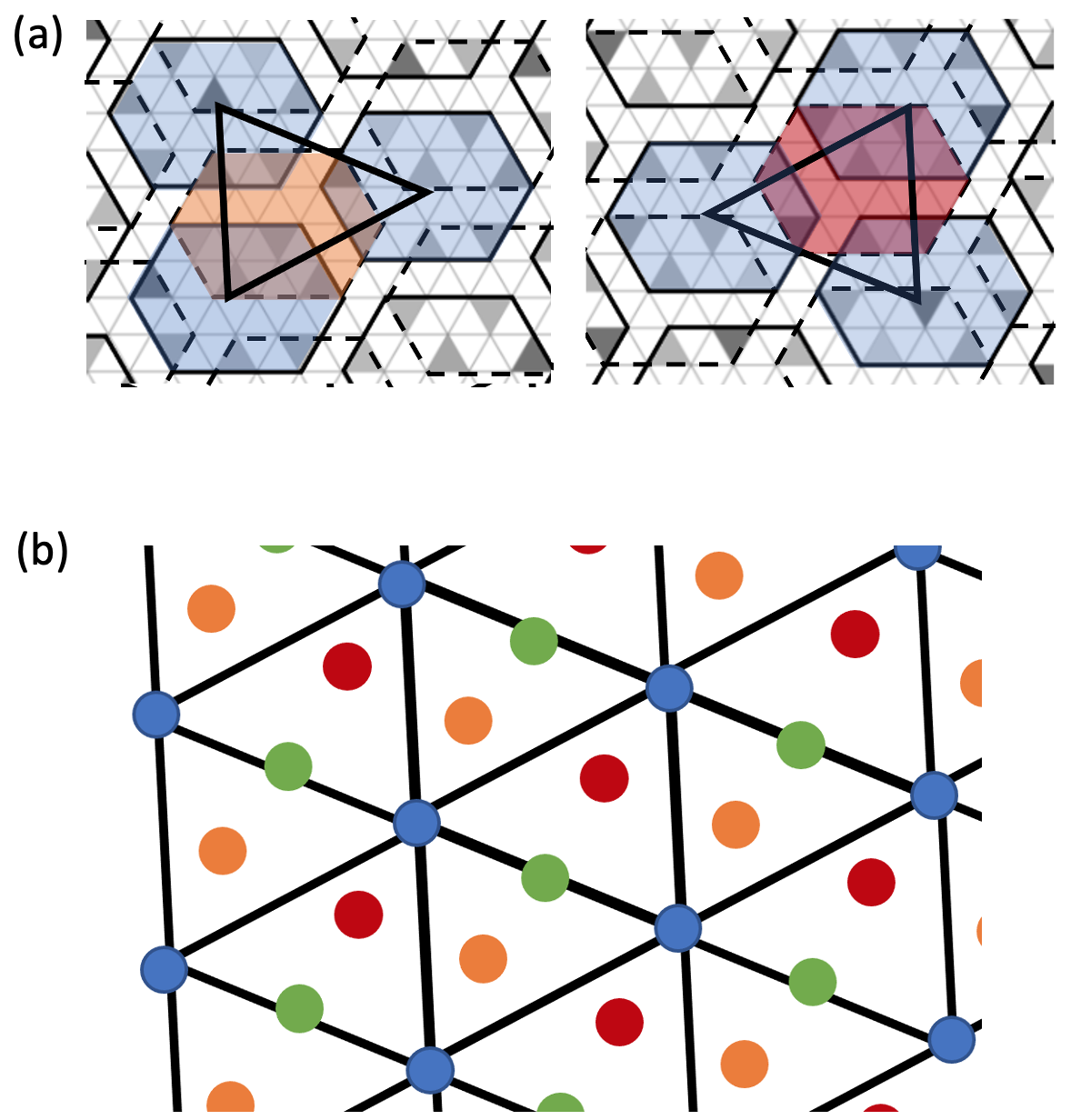}
    \caption{The resonance in QLS and effective lattice model. a) The additional ``flippable" configurations in QLS. Left: an additional ``down"-pointing configuration (orange shade) emerges by arranging three ``up"-pointing configurations (blue shade) arranged in a right-pointing triangle (thick black lines). Right: an additional ``up"-pointing configuration (red shade) emerges by arranging three ``down"-pointing configurations (shaded in blue) arranged in a left-pointing triangle (thick black lines). b) The lattice constructed by viewing the centers of the ``flippable" configurations in a) (marked by the elongated hexagons) as lattice sites. The blue, orange, and red sites correspond to the shaded plaquettes with the same colors in a). (The green sites are not shown explicitly in a).  }
    \label{fig:QLS_melt}
\end{figure}

Here we propose a minimal effective model for studying the competition between QLS and nearby QLL phases by regarding the supercells of the ``flippable" configurations as local degrees of freedom. How these local degrees of freedom are related to the QLS is illustrated in \autoref{fig:QLS_melt} (a). The effective model is defined on a distorted triangular lattice with four orbitals per unit cell labeled by the four colors in \autoref{fig:QLS_melt} (b). For each site, we define a local three-dimensional Hilbert space spanned by $|0\rangle, |1\rangle, |-1\rangle$. The two states $|\pm 1\rangle$ denote the two ``flippable" configurations of the lemniscate operator, where $|1\rangle$($|-1\rangle$) denotes the orientation of the middle two triangles being ``up"(``down")-pointing (\autoref{fig:QLS_melt} (a)). The $|0\rangle$ state denotes an ``un-flippable" configuration. By considering the possible resonance within the QLS, we arrive at the minimal effective Hamiltonian that could describe the melting of QLS to a nearby QLL phase,
\begin{equation}
    H_{\textnormal{eff}}^{\textnormal{QLS}} = - J_Q \sum_{\langle i,j\rangle,a }\tau^z_{i,a} \tau^z_{j,a} - \tilde{t} \mathcal{P}\left(\sum_{i,a} \tau^x_{i,a}\right) \mathcal{P}, 
\end{equation}
where $a \in \{1,2,3,4\}$ labels the orbitals,  $i,j$ label the sites, $\tau^{x,y,z}$ are the Pauli matrices acting on $|\pm 1\rangle $ states and $\mathcal{P}$ is a projection operator. $J_Q > 0 (<0)$  denotes the nearest neighbor FM (AFM) coupling between the same orbitals that favors a QLS state. $\tilde{t}$ denotes the strength of the lemniscate resonance. Without the projector $\mathcal{P}$, the Hamiltonian $H_{\textnormal{eff}}^{\textnormal{QLS}}$ is the same as a transverse field Ising model, whose phase diagram is well-known. Now let us specify the definition of $\mathcal{P}$, which encodes the non-trivial correlation depicted in \autoref{fig:QLS_melt} (a). The projection imposes local constraints such that for each triangle formed by the three nearest neighbor sites with the same colors in \autoref{fig:QLS_melt}, if the three sites are all in state $|1\rangle$, the site that is enclosed by the triangle has to be in state $|-1\rangle$ ($|0\rangle$) for right (left)-pointing triangles, similar for the case where all the three sites are in state $|-1\rangle$. The projection $\mathcal{P}$ makes $H_{\textnormal{eff}}^{\textnormal{QLS}}$ non-trivial and whether the ground state of $H_{\textnormal{eff}}^{\textnormal{QLS}}$ could be a spin liquid is an open question.

%{\color{purple}
\section{E. Emergent Luttinger liquid in brick-wall phase upon doping} \label{sec:LL}
\subsection{Stability of the brick-wall phase}
If there is no quantum fluctuation given by the hopping of the electrons, the brick-wall phase is the ground state of the Hamiltonian for $U_4 < 0$. We find the brick-wall phase to be stable under small perturbations of quantum fluctuation effects, due to the high degree of hopping necessary for the lemniscate resonance. 

Specifically, the energy gain from quantum fluctuation is given by the energy scale of the lowest-order ring exchange operator, i.e. the ``lemniscate" operator, which is $\sim t^8/U^7$. And therefore, as long as $t \ll |U_4| (U/|U_4|)^{7/8}$, the brick-wall phase is favored. Since $U \gg |U_4|$, the brick-wall phase is stable even when $t > |U_4|$. And in the following, we will see that this ``excess" kinetic energy can give rise to non-trivial Luttinger liquid-like behavior when we dope away from 1/3 filling.

\subsection{Subextensive degeneracy}
\begin{figure}
    \centering
    \includegraphics[width=0.48\textwidth]{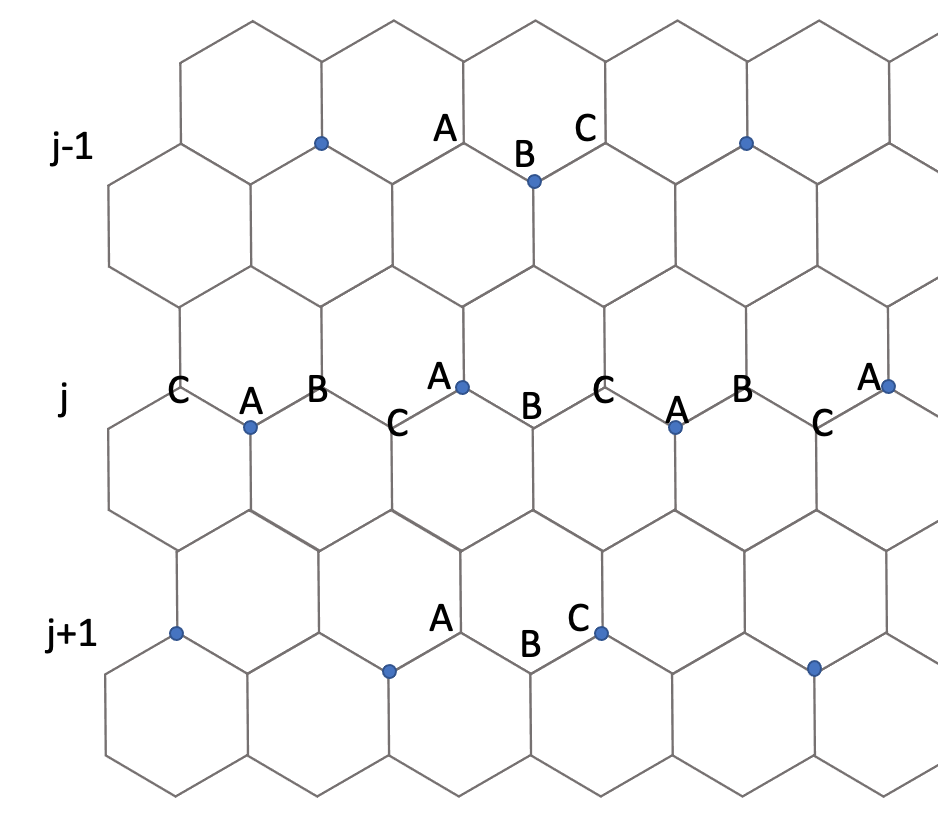}
    \caption{Brick-wall phase. $j$ labels different rows. On each row, A, B, and C label different sublattices.}
    \label{fig:brick-wall}
\end{figure}

 The brick-wall phase can be viewed as an array of 1D stripes as in Fig.\ref{fig:brick-wall}. On each row, the charges can sit either on $A$, $B$, or $C$ sublattices. As illustrated in Fig.\ref{fig:brick-wall}, if the charges of the nearby chains are located at different sub-lattice sites, there are additional 4-th nearest neighbor bonds. To simplify the notation, we label the different phases of the stripes on each as $a$, $b$, and $c$ by whether their charges sit on sites $A$, $B$, or $C$. Starting from the zeroth row, we can build the brick-wall state one row at a time. The attractive $U_4$ interaction favors the phase of row $j+1$ to be different than row $j$, say if the row $j$ is in phase $a$ then the row $j+1$ can be in either phase $b$ or phase $c$. And therefore, the degeneracy of the brick-wall phase is $3\times 3\times 2^{L-1}$, where $L$ is the number of rows and two factors of $3$ denote the phase of the zeroth row and three different orientations of the stripes.

\subsection{Single-chain model for emergent Luttinger liquid with $K=1/3$}
\begin{figure}
    \centering
\includegraphics[width=0.48\textwidth]{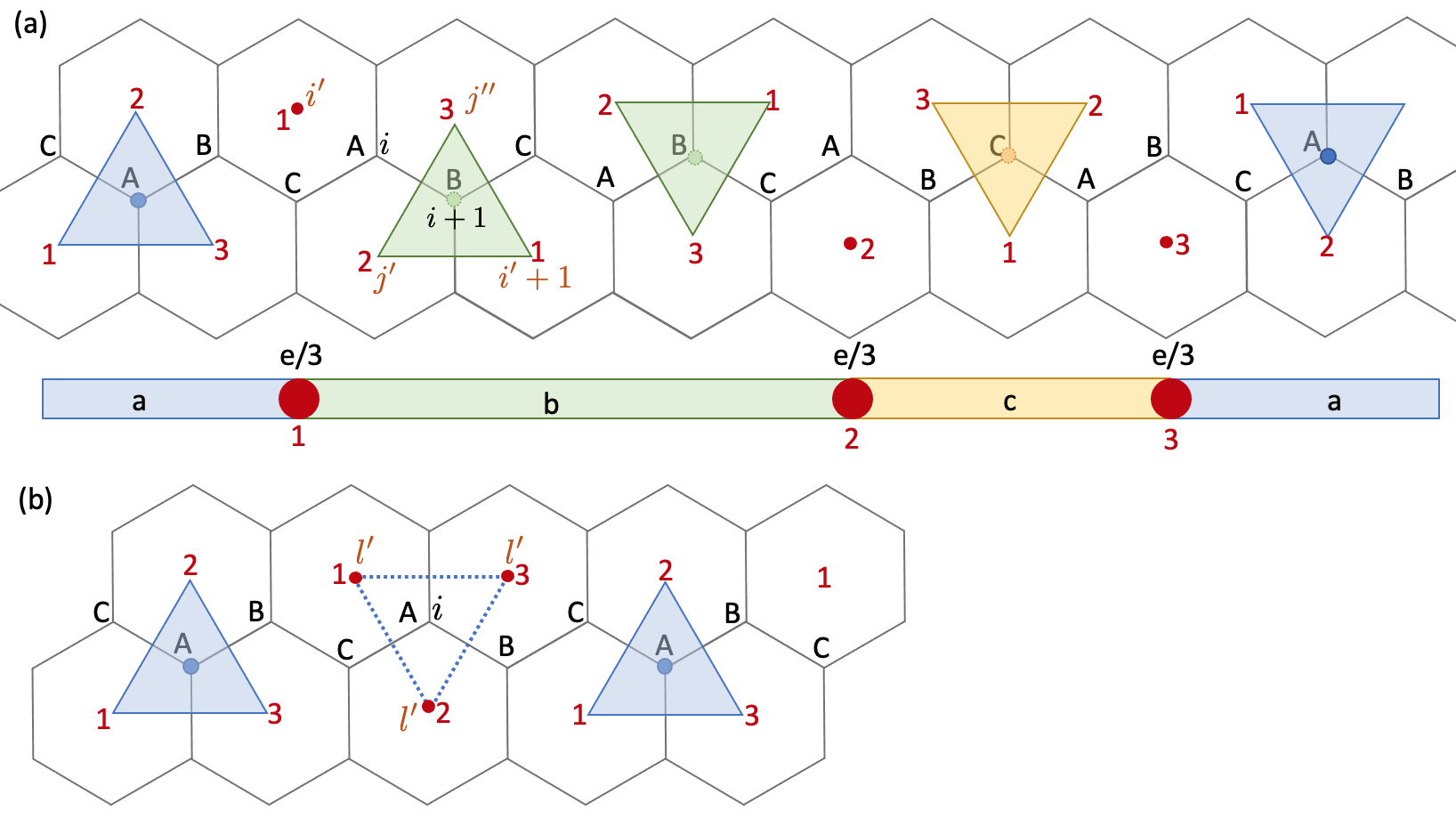}
    \caption{Domain walls between different orderings. (a) The upper panel shows the original lattice, with ``A", ``B", ``C" denoting the sites of the physical electrons and the triangle denotes the mapping to trimers. The original lattice is defined on the zig-zag chain of the honeycomb lattice and the ``dual" lattice is defined on the centers of the plaquettes of the honeycomb lattice. The bottom panel is a cartoon picture depicting different charge ordering phases (labeled by a,b, and c) and the domain walls. Different flavors of domain wall excitations are labeled by the number of plaquettes (red numbers of the upper panel). For example, solitons of flavor ``1" labels the domain wall between phase ``a" and ``b", etc. The hopping process only moves the solitons of the same type and the soliton can hop to its nearest neighbor (with same flavor) only when there is no soliton in between. $i$, $i+1$ label the original lattice and $i'$, $i'+1$, $j'$, $j''$ label the ``dual" lattice as in Eq.\ref{eq:H_eff_1d}. (b) Mapping between physical electron and solitons, where $i$ labels the site where an electron is annihilated (denoted by the dotted triangles) and the $l'$ labels the sites on the ``dual" lattice where the solitons are created. }
    \label{fig:domain-wall}
\end{figure}

Now, let us consider small hole-doping to the Brick-wall phase. 
In the following, we will consider a single isolated chain as shown in Fig.\ref{fig:domain-wall}, ignoring hopping and interaction between chains. 
Since the motion of a single hole is confined within 1d and the hole density is dilute, we expect the single-chain approximation to work well. 

Curiously, as opposed to the conventional commensurate-incommensurate (C-IC) transition of doping a Mott insulators near 1/3 filling, where the Luttinger parameter is expected to be $1/9$ (see Ref.\cite{giamarchi2003quantum} Ch.4 and the reference therein), we find the emergent Luttinger liquid in the single-chain model to have Luttinger parameter $1/3$, which is a consequence of the strong cluster charging interaction, as we elaborate in the following.

In the single-chain approximation, the effective 1d Hamiltonian can be written as,
\begin{equation}
\begin{split}
    H_{1d} &= t \sum_{\langle ij \rangle} f_i^\dag f_j + h.c.\\
    &- U_4 \sum_{\langle i,j\rangle_4} n_i n_j \\&+ 3U \sum_i n_i^2 + 4U \sum_{\langle i,j\rangle}n_i n_j + 2U \sum_{\langle i,j\rangle_2}n_i n_j,
\end{split}
\label{eq:H_1d}
\end{equation}
where the first term is the kinetic energy with $f_i^\dagger$ denoting the operator for an electron creation at site $i$ restricted to a chain (see Fig.\ref{fig:domain-wall} for the geometry of the chain), the second term and the third term are the fourth-nearest-neighbor density-density interaction,  and the cluster-charging interaction is restricted to the 1d chain, respectively.

Now, we turn to the low-energy description of the Hamiltonian Eq.\ref{eq:H_1d} in terms of the soliton representation of the domain walls \cite{giamarchi2003quantum}. Since we are interested in the regime where $U \gg t, U_4$, we cannot apply the usual bosonization that treats the interaction perturbatively in Eq.\ref{eq:H_1d}. Instead, we start from the strong coupling regime where $U\rightarrow \infty$ and project out all the configurations that are prohibited by the cluster charging to arrive at effective kinetic energy, which has a complicated form but can be further simplified using the soliton representation,
\begin{equation}
    \begin{split}
        &H_{\text{eff},K} \\
        &= t \sum_{i} Q_{i,i+1} (1-n_{i-2})(1-n_{i-1})(1-n_{i+2})(1-n_{i+3})\\
        &= -t \sum_{i',\alpha} \tilde{Q}_{\alpha,i',i'+1}(1-\tilde{n}_{\alpha+1,j'})(1-\tilde{n}_{\alpha+2,j''}),
    \end{split}
    \label{eq:H_eff_1d}
\end{equation}
where the first line is written in terms of the physical fermions $f_i$, $Q_{i,i+1} \equiv f_i^\dag f_{i+1}+f_{i+1}^\dag f_{i}$ and the second line is written in terms of the ``soliton" fields $d_{i'}$, $\tilde{Q}_{\alpha, i',i'+1} \equiv d_{\alpha,i'}^\dag d_{\alpha, i'+1}+d_{\alpha,i'+1}^\dag d_{\alpha,i'}$ and $\tilde{n}_{\alpha,i'} = d_{\alpha,i'}^\dag d_{\alpha, i'}$. We define the ``soliton" field  $d_{\alpha, i'}^\dag$ to be the domain-wall creation operator for flavor $\alpha$ at site $i'$ defined on the ``dual" lattice (labeled by the red numbers in Fig.\ref{fig:domain-wall} a) and $\alpha\in \{1,2,3\}$ labeled the flavor of the solitons. The $j'$ and $j''$ in the last line of Eq.\ref{eq:H_eff_1d} label the sites between $i'$ and $i'+1$ for the soliton of flavor $\alpha$.(Fig.\ref{fig:domain-wall}a) The solitons of different flavors do not mix each other since the nearest-neighbor hopping only moves the defects of the same type (Fig.\ref{fig:domain-wall} a), and therefore the flavor index is a  good quantum number at low energy. Note that the mapping between the Hilbert space in the number basis of the physical electrons and that of the solitons is one-to-one. However, the mapping between operators is non-trivial. Since annihilating one physical electron creates three domain walls, we have $(1-n_{i-2})(1-n_{i-1}) c_i (1- n_{i+1}) (1- n_{i+2}) = (d_1^\dag d_2^\dag d_3^\dag)_{l'}$, where $l'$ denotes the three plaquettes surrounding site $i$ (Fig.\ref{fig:domain-wall}b). 

It is readily seen that from the strong coupling expansion, the ``soliton" fields gain emergent ``flavor" d.o.f, as opposed to the solitons in the C-IC transition, which ultimately gives rise to different Luttinger parameters, as we explain in the following.

First, let us briefly review the results in the C-IC transition. 
For the C-IC transition tuned by doping (Mott-$\delta$ transition), the Luttinger parameter is shown to be $1/n^2$ for spinless fermions \cite{giamarchi2003quantum} in the infinitesimal doping limit, where $n$ is the commensurability, $n=3$ for 1/3-filling. One way to derive this Luttinger parameter is to relate the correlation function of the physical fermion to the solitons. Applying bosonization, the fermion density $\rho(r) = \frac{1}{\pi} \nabla \phi(r)$. Since the density of the solitons is three times larger than that of the fermions for 1/3-filling, we have the soliton field $\tilde{\phi} = 3 \phi$. Now let us consider the correlation function $\langle e^{i 2\phi(r)} e^{-i 2\phi(0)} \rangle$. On one hand, it is $\sim x^{-2K}$, where we ignore all the non-universal factors and $K$ is the Luttinger parameter of the LL written in $\phi$ fields. On the other hand, $\langle e^{i 2\phi(r)} e^{-i 2\phi(0)} \rangle = \langle e^{i 2\tilde{\phi}(r)/3} e^{-i 2\tilde{\phi}(0)/3} \rangle \sim x^{-2\frac{1}{9}}$ since $\tilde{\phi}$ is free in the infinitesimal doping limit. Hence $K = \frac19$ for the Mott-$\delta$ transition.

Now let us consider the emergent LL in the brick-wall phase. Since we have three independent flavors, we define the soliton fields $\tilde{\phi}_i$ ($i = 1,2,3$) and $\tilde{\phi}_1+ \tilde{\phi}_2+ \tilde{\phi}_3 = 3 \phi$ due to the relation between the density of the physical fermions and the solitons. Analogously, we can compute the correlation function $\langle e^{i 2\phi(r)} e^{-i 2\phi(0)} \rangle$.
In the dilute limit, the soliton density $\tilde{n} \rightarrow 0$. From Eq.\ref{eq:H_eff_1d} it is readily seen that Hamiltonian becomes quadratic, and hence the solitons are free. Therefore $\langle e^{i 2\phi(r)} e^{-i 2\phi(0)} \rangle = \Pi_{\alpha=1}^3 \langle e^{i 2\tilde{\phi}_\alpha (r)/3} e^{-i 2\tilde{\phi}_\alpha (0)/3} \rangle \sim x^{-2\frac{1}{3}}$ and $K = \frac13$. For finite but small doping and for finite $U_4$, we expect the emergent Luttinger liquid behavior still holds, but the Luttinger parameter $K$ is tuned by doping and the interaction $U_4$, which we leave for future study.

The emergent Luttinger liquid behavior has unique experimental signatures \cite{Kane1992Phys.Rev.Lett., Kane1992Phys.Rev.B}. First of all, the tunneling DOS $\rho(\epsilon - \epsilon_F) \sim (\epsilon - \epsilon_F)^{\alpha}$ \cite{Kane1992Phys.Rev.Lett., Kane1992Phys.Rev.B}, where $\alpha = \frac12 (K+K^{-1} -2)$, assuming spin-valley polarization, which can be observed in STM experiments. Secondly, Ref.\cite{Georges2000Phys.Rev.B} argues that for weakly coupled LLs with inter-chain hopping $t_\perp$ and intra-chain hopping $t$, the inter-chain conductance $G$ has scaling behavior $\sim \omega^{2 \alpha -1}$ in the regime when the temperature is higher than a crossover energy scale $E^* \sim t_\perp (t_\perp/t)^{\alpha/(1-\alpha)}$. At a relatively low temperature when $E^*< k_B T \ll eV$, $G = dI/dV \sim V^{2 \alpha-1}$ deviated from Ohmic behavior. At a higher temperature when $k_B T \gg eV$, $G \sim T^{2 \alpha-1}$. Thirdly, we expect a violation of the Wiedmann-Franz Law. Since for small doping $K\sim 1/3$, the conductance $G$ approaches zero at zero temperature due to Anderson localization in the presence of impurity and therefore the Lorentz number $L = \kappa /TG$ has divergent behavior at low-temperature \cite{Kane1996Phys.Rev.Lett.a, Li2002Europhys.Lett.}.

\section{F. Emergent 1-form symmetry, restricted mobility, and sub-extensive ground state degeneracy}
\subsection{Emergent 1-form symmetry}
In this section, we explain in more detail that the cluster charging constraints lead to emergent $U(1)\times U(1)$ 1-form symmetry at low energy.

To construct emergent symmetry generators, we first write down the low-energy effective Hamiltonian $H_{\text{eff}}$, including all the local operators that commute with the cluster charging terms. Then we identify the emergent symmetry generators as operators that commute with all the terms in $H_{\text{eff}}$ \cite{Pace2023}. 

\begin{figure}
    \centering
    \includegraphics[width = 0.45\textwidth]{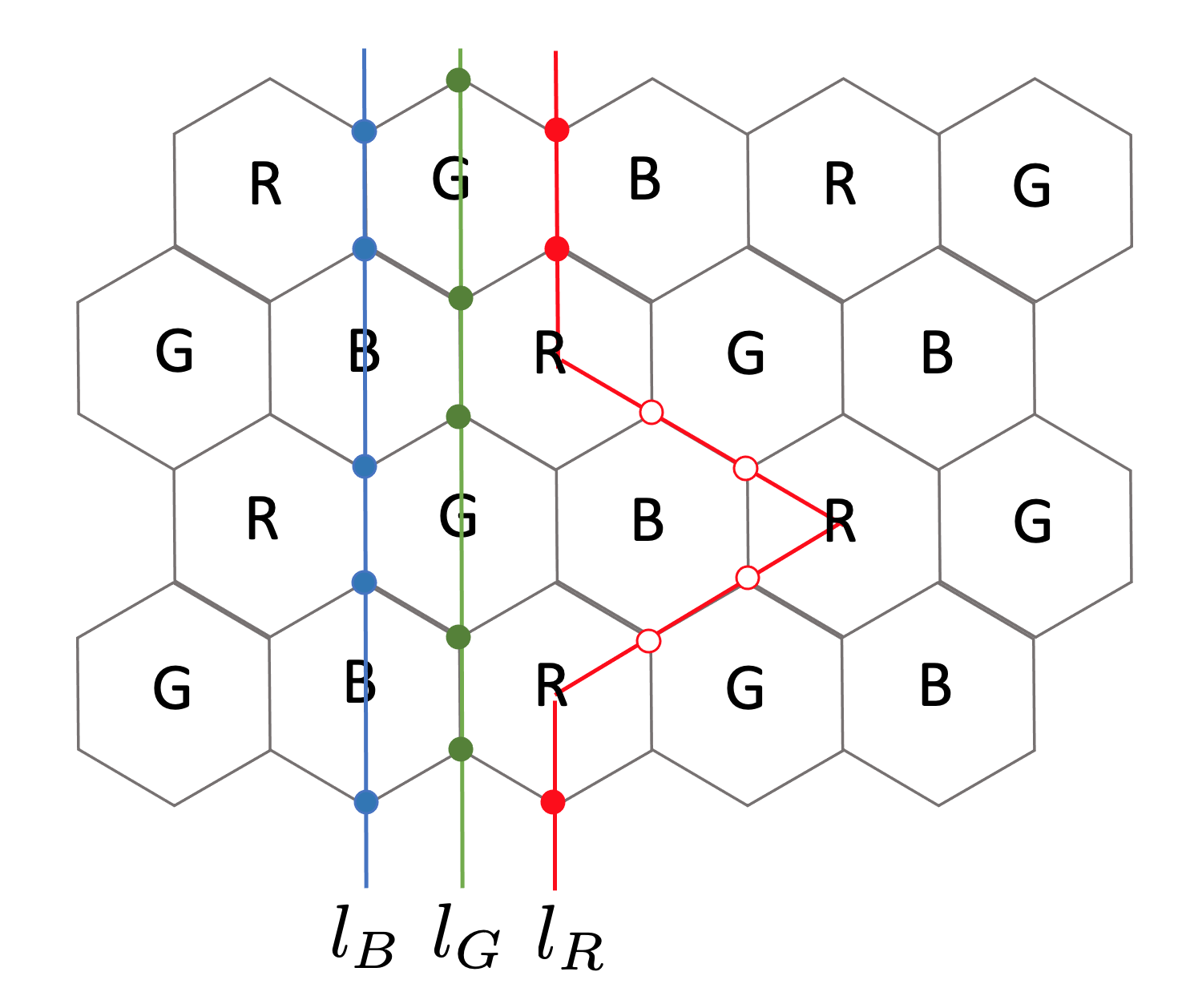}
    \caption{Emergent 1-form symmetry. R,G, and B label different honeycomb plaquettes. The symmetry generators are defined along the lines that only pass the plaquettes of the same type. The filled (unfilled) circles denote adding (subtracting) the density operators on the corresponding sites.}
    \label{fig:1_form}
\end{figure}

Since all the operators due to quantum fluctuation can be built from the ring exchange operator $\mathcal{O}$ on a single plaquette, one can verify that there are emergent 1-form symmetries generated by the string operators illustrated in Fig.\ref{fig:1_form}. To define the 1-form symmetry generators, we first tri-partite the honeycomb plaquette into R, G, and B types. For each type of honeycomb plaquette, we can draw an arbitrary string that passes the centers of the plaquettes. The symmetry generator has support on the sites that intersect with the string, as denoted by the circles in Fig.\ref{fig:1_form}. We can then define three $U(1)$ 1-form symmetry generators as $Q_{\alpha} = \sum_{i \in l_{\alpha}} (-1)^{S(i)} n_i$, where $\alpha \in \{R,G,B\}$ denotes the type of the string, $l_{\alpha}$ denotes all the sites on the honeycomb lattices that intersects with the string and $S(i) = \pm 1$ depending on the direction of the string. The rule is that if the string passes the same sublattice, there is a sign flip. (See the red string in Fig.\ref{fig:1_form} for an example.) We also note that although we can define three charges, only two of them are independent. To see this, if we put the system on a torus, the summation of the three different charge operators, if supported on three topologically equivalent non-contractible loops, can be decomposed into the sum over charges around many honeycomb plaquettes, whose values are given by the cluster charging condition. Therefore, the emergent 1-form symmetry is only $U(1) \times U(1)$.

Hence, if we put the system on a torus, there will be $O(L^4)$ topological sectors, where $L$ is the linear system size and the power comes from two $U(1)$ symmetries along two topologically distinct non-contractible loops of the torus.

\subsection{Charged string operator and restricted mobility of the fractionally charged excitations}
Now let us consider doping a single hole. The motion of the defects is determined by products of local operators that can move the defects while still respecting the cluster charging condition.\footnote{Here, the defects are defined w.r.t to honeycomb plaquette, i.e., a defect plaquette means that it has zero electrons around its six sites, and we can also label the defects by R, G, B, depending on the plaquette it lives on. Note that a single physical hole creates one defect of R, G, B types each.} As we will see in the following, the operators that move the defects are ultimately related to the charged string operators of the emergent 1-form symmetry.

\begin{figure}
    \centering
    \includegraphics[width = 0.45\textwidth]{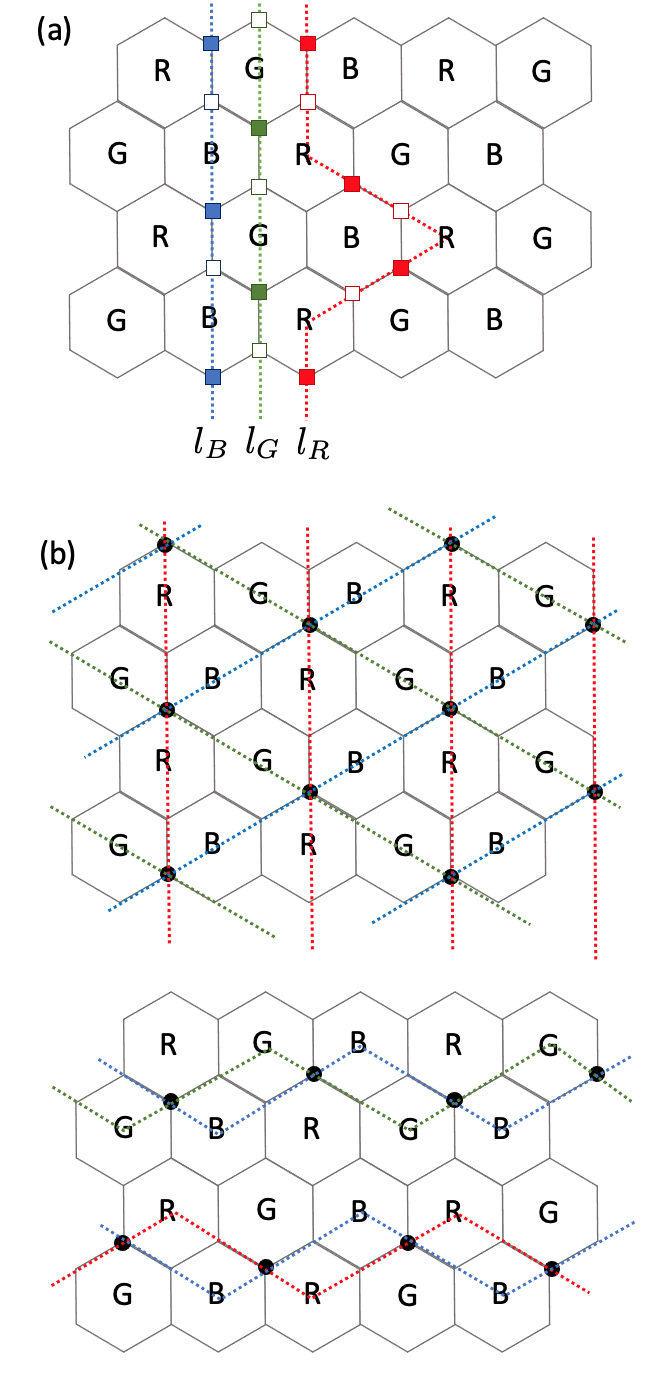}
    \caption{(a) Charged operators of the $U(1)\times U(1)$ 1-form symmetry. The solid/hollow square denotes fermion creation/annihilation operators, and the operator is a product of these fermion operators along the string. (b) ``Compatible" string operators in $\sqrt{3}\times\sqrt{3}$ (top) and brick-wall (bottom), which give rise to the motion of the defects of the same type as the color of the string.}
    \label{fig:charged_strings}
\end{figure}

Let us first specify all the charged operators under the $U(1)\times U(1)$ 1-form symmetry. Those operators are string operators given by products of fermion creation/annihilation operators (see Fig. \ref{fig:charged_strings}a) and the supports of these strings have the same structure as the 1-form symmetry generators. There are three types of string operators, w.r.t the tri-partition of the honeycomb plaquettes. If we consider a finite segment of these strings, it moves a defect from one end to the other end, without creating more defects along the way. 
Moreover, the different types of string operators correspond to moving different types of defects, depending on the color of the strings. 

There is a caveat to the above statement. Besides moving the defects, some string operators annihilate a specific state. We call these string operators ``incompatible" with the state. Those string operators that act non-trivially on a state are called ``compatible". But since there are also local ring-exchange operators that do not create defects, instead of considering one single state $|\psi\rangle$, we should include all the states that are connected to $|\psi\rangle$ by some local operator $\hat{L}$. To be more precise, a string operator $\hat{S}$ is said to be compatible with $|\psi\rangle$ if there exists a local operator $\hat{L}$ (that commutes with the cluster-charging), such that $\hat{S} \hat{L}|\psi \rangle \neq 0$. And only the compatible string operators contribute to the motion of a single defect of the corresponding state.

For the $\sqrt{3}\times \sqrt{3}$ and brick-wall phases, the only compatible string operators are rigid, namely their shapes cannot deform (see Fig.\ref{fig:charged_strings}b), which constrains the motion of the solitons to be 1d, i.e., lineon-like. However, in the phases dominated by quantum fluctuations through a local resonance pattern, for example, the QLS/QLL phases, there is no restriction on the mobility of the 1/3 charged excitations. \footnote{We would like to distinguish restricted mobility from confinement. Since both terminologies appear in the manuscript, it may cause some confusion. The restriction of the motion is associated with the cluster charging energy scale while the confinement of the excitations, for example, the solitons in $\sqrt{3}\times \sqrt{3}$ phases, is associated with energy scales at low energy, namely the kinetic energy and further neighbor interactions.}

\subsection{Sub-extensive ground state degeneracy and hidden symmetry}
Now we return to the sub-extensive ground state degeneracy in the brick-wall phase specified in SM sec. E. First of all, the degeneracy is not projected by topology. If we add further range density-density interaction, for example, thirteenth nearest neighbor, the sub-extensive ground state degeneracy will be lifted and result in discrete degeneracy. And therefore, this sub-extensive degeneracy is accidental and there is a hidden symmetry in the brick-wall phase explained in the following.

\begin{figure}
    \centering
    \includegraphics[width=0.45\textwidth]{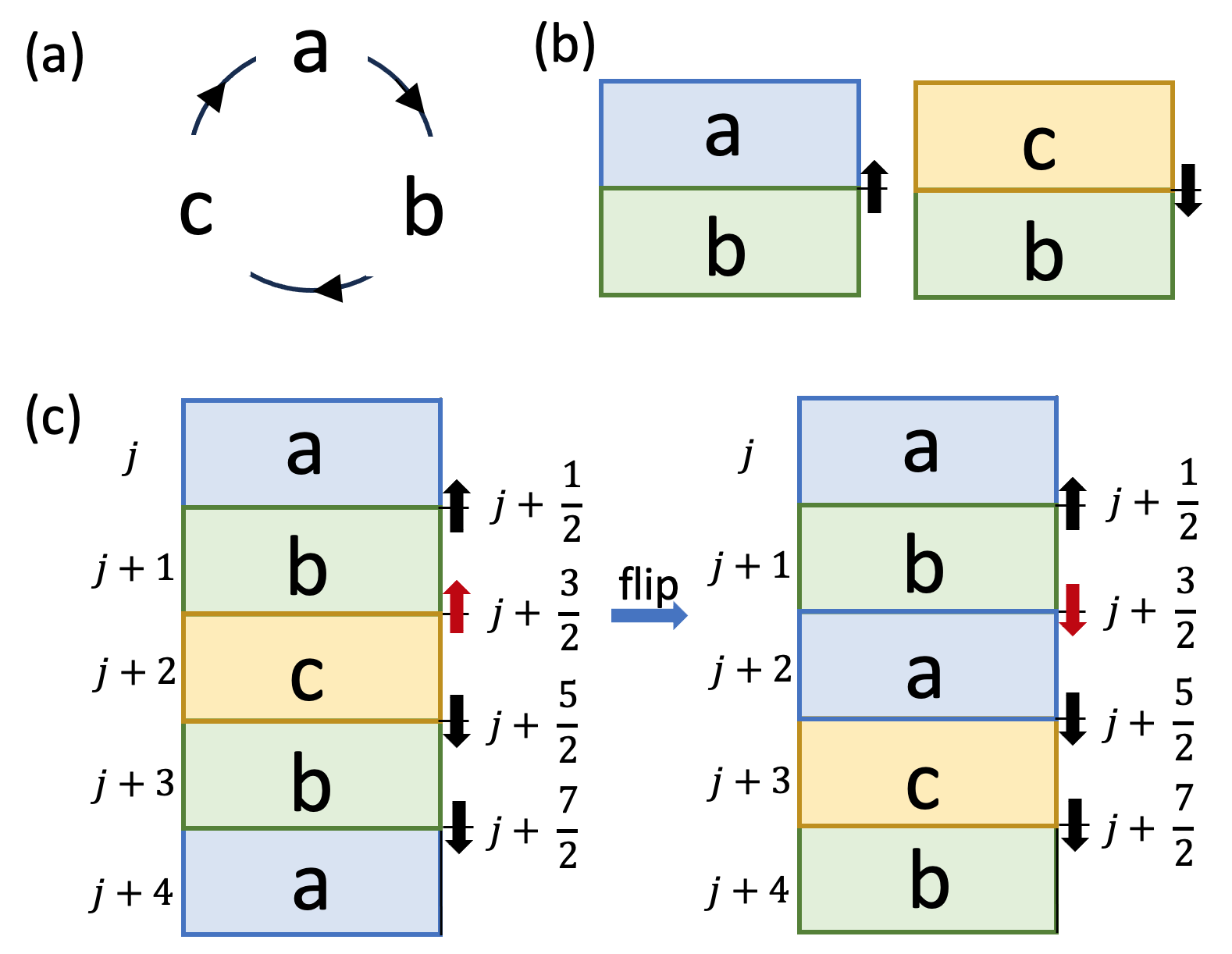}
    \caption{Mapping from brick-wall configuration to Ising chain. (a) Cyclic permutation of the three-sub lattice ordering. (phase a, b, c following the convention in Fig.\ref{fig:domain-wall}) (b) Brick-wall configuration maps to Ising spin configurations. The corresponding spin d.o.f is defined on the dual lattice between the two rows. (c) Single spin flip in the Ising chain maps to non-local shifts in the brick-wall phase, i.e., flipping a spin $\uparrow$ ($\downarrow$) to a spin $\downarrow$ ($\uparrow$) maps to shifting the phase clockwise (counter-clockwise) for the rows below the flipped spin and onward. In the example, $j, ..., j+4$ label the rows of the brick-wall configuration, and $j+1/2,...,j+7/2$ label the sites of the dual-spin chain. The spin on-site $j+3/2$ (marked red) is flipped from $\uparrow$ to $\downarrow$, which maps to the change in brick-wall configurations, with all the configurations of the rows from $j+2$ onward shift clockwise according to the rules in (a).  }
    \label{fig:bw-sub-deg}
\end{figure}

To define the generator of the hidden symmetry in the brick-wall phase, it is convenient to map the brick-wall configuration to a 1d Ising spin chain. As pointed out in SM sec. E, nearby rows have to be of different sub-lattice ordering chosen from phases a, b, c. Hence, we can label the difference of the sub-lattice ordering between row $j$ and row $j+1$ by an Ising degree of freedom. The rule is that we first define a cyclic permutation as $a\rightarrow b \rightarrow c \rightarrow a$ (Fig.\ref{fig:bw-sub-deg} a) and if the phases of row $j\rightarrow$ row $j+1$ agree with the permutation, we assign spin $\uparrow$ on the sites of the dual lattice perpendicular to the rows and vice versa (Fig. \ref{fig:bw-sub-deg} b). For each site on the dual lattice, the spin can be either $\uparrow$ or $\downarrow$, which maps to the different brick-wall configurations. The hidden symmetry is, therefore, the spin-flip on the Ising spin chain, which maps to non-local shifts in the brick-wall phase (Fig. \ref{fig:bw-sub-deg} c) and are related to the products of the non-trivial, compatible string operators in Fig.\ref{fig:charged_strings} (b). Analogous to the 0-form symmetry (global symmetry), if the ground states are degenerate and the symmetry generator maps one ground state to another, it is in the spontaneous symmetry breaking (SSB) phase. Since the generators of the hidden symmetry toggle between different brick-wall configurations, the sub-extensive degeneracy can be attributed to the spontaneous breaking of the hidden symmetry.

\section{G. Other $n/3$ filling fractions and experimental implications}
In the main text, we consider $-11/3$ filling (counting from neutrality) and we discuss some other $n/3$ filling fractions in this section. In particular, we will consider two and three electrons per unit cell, corresponding to filling $-10/3$ and $-3$. Note that four electrons per unit cell ($-8/3$ filling) and five electrons per unit cell ($-7/3$ filling) can be related to $-10/3$ and $-11/3$ fillings by considering holes instead of electrons starting from the Mott insulator with one electron per site. We will also discuss scenarios with higher doping (greater than six electrons per plaquette) such that there is double occupancy.

We start with $-10/3$ filling. For $-11/3$ filling the leading order correction to the cluster charging interaction in terms of density-density interaction is 4-th nearest neighbor since the electrons are at least 4-th nearest neighbor apart from each other. Now if we consider two electrons per plaquette, we need to take into account the imbalance of the 1st, 2nd, and third nearest neighbor interaction due to the finite extent of the Wannier orbitals. And therefore, the interaction Hamiltonian up to third-nearest neighbor terms can be written as,
\begin{equation}
    \begin{split}
        H_{\text{int}} =& V_0 \sum_i (n_i)^2 \\
        &+ \sum_{m = 1}^{3} \sum_{\langle ij \rangle_{m}} \left( V_{m} n_i n_j + J_m c_{i,\alpha,\tau}^\dag c_{j,\alpha,\tau} c_{j,\beta,\tau'}^\dag c_{i,\beta,\tau'} \right),
    \end{split}
    \label{eq:H_int_3}
\end{equation}
where we assume $SU(4)$ spin-valley symmetry, which is a good approximation for MATBG, and we expect the effect of $SU(4)$ symmetry breaking to be more prominent away from the magic angle. 

\begin{figure}
    \centering
    \includegraphics[width = 0.45\textwidth]{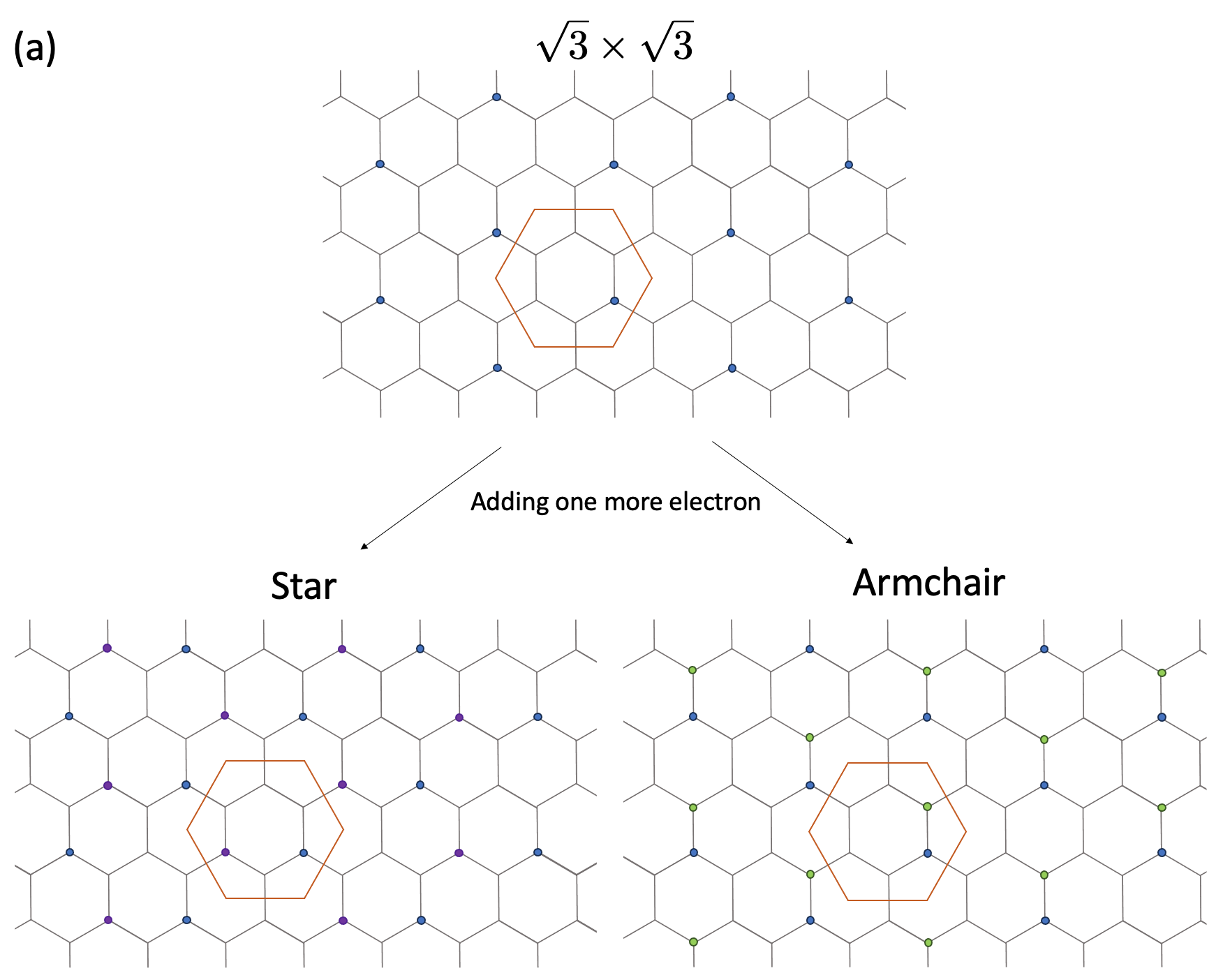}

    \includegraphics[width = 0.45\textwidth]{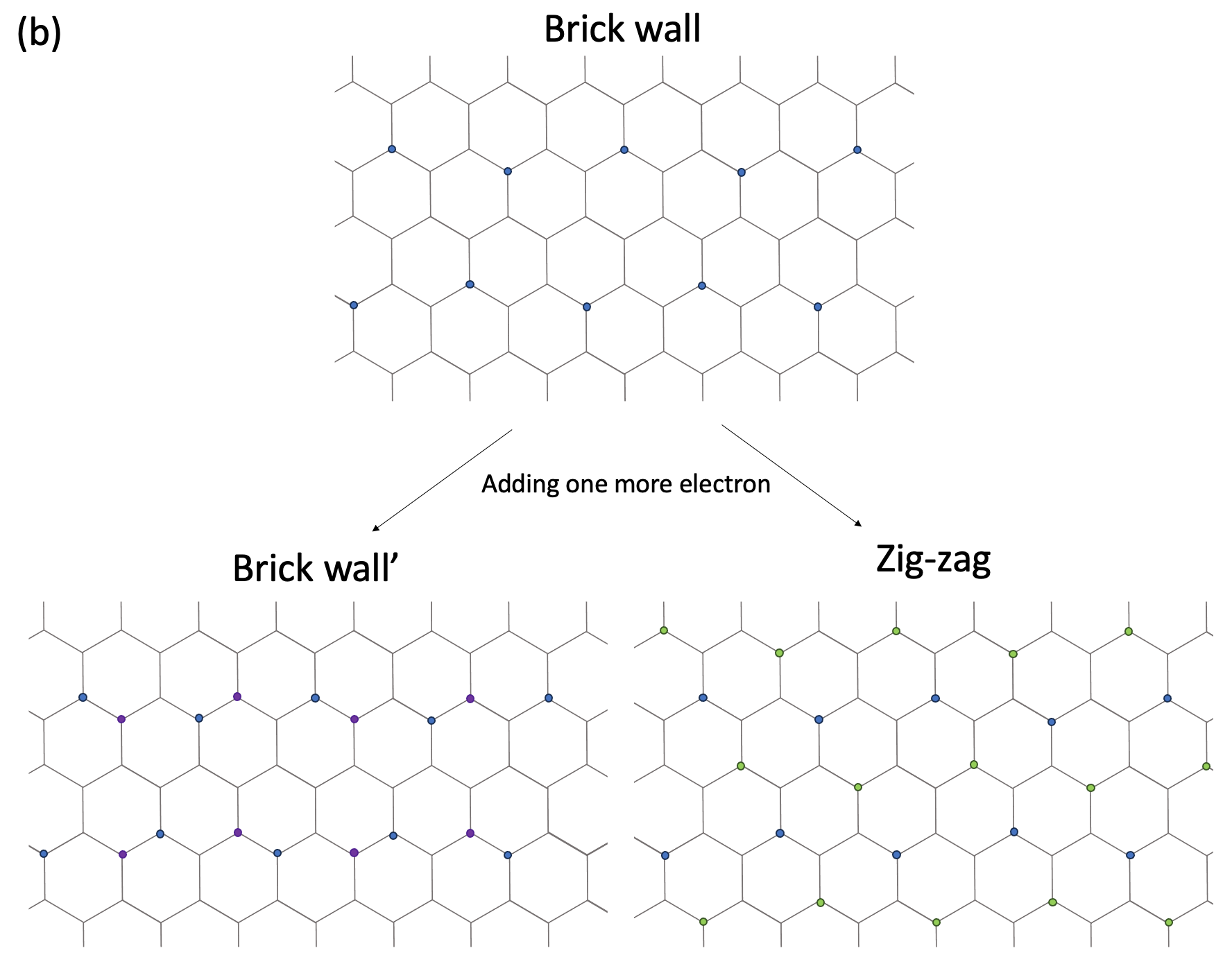}
    \caption{Relationship between the phases at $-11/3$ filling and $-10/3$ filling. (a) Adding one electron per plaquette to the $\sqrt{3} \times \sqrt{3}$ phase gives rise to star or armchair phases depending on whether the additional electrons sit at the same sublattices or different ones. The orange hexagon denotes the unit cell. (b) Adding one electron per plaquette to the brick-wall phase gives rise to brick-wall or zig-zag phases depending on whether the additional electron sits along the ``bricks" or between the ``bricks". }
    \label{fig:four_phases}
\end{figure}

By considering the above Hamiltonian, we find four competing orderings, dubbed as ``star", ``armchair", ``brick-wall' " and ``zig-zag" (see Fig.\ref{fig:four_phases}). Due to the spin-valley ferromagnetic exchange, these phases are all spin-valley ferromagnets. The star and armchair phases can be viewed as descendants from the $\sqrt{3}\times\sqrt{3}$ phase of $-11/3$ filling by adding one more electrons per plaquette. Likewise, the brick wall and zig-zag phases are derived from the brick-wall phase, and they all have sub-extensive degeneracies.

The energy densities of these phases are,
\begin{equation}
    \begin{split}
        E_{\text{star}} &= 3 U_2 \\
        E_{\text{armchair}} &= U_1 + U_3\\
        E_{\text{brickwall}} &= U_1 + U_2\\
        E_{\text{zigzag}} &= 2 U_2 + U_3.
    \end{split}
\end{equation}
where $U_i = V_i-J_i/2$. It is readily seen that in the strong coupling regime, the phase diagram is more complicated than that of $-11/3$ filling is summarized in Table. \ref{tab:energy_four_phases}.
\begin{table}[h]
    \centering
    \begin{tabular}{|c|c|c|}
    \hline
         & $U_2 - U_3 > 0$  & $U_2 - U_3 < 0$\\
         \hline
       $U_1-2U_2 > 0$  & Zig-zag & Star\\
       \hline
       $U_1-2U_2 < 0$ & Armchair & Brick-wall' (Strip)\\
       \hline
    \end{tabular}
    \caption{Competition between the four phases at $-10/3$ filling and $-1$ filling.}
    \label{tab:energy_four_phases}
\end{table}

\begin{figure}
    \centering
    \includegraphics[width = 0.25\textwidth]{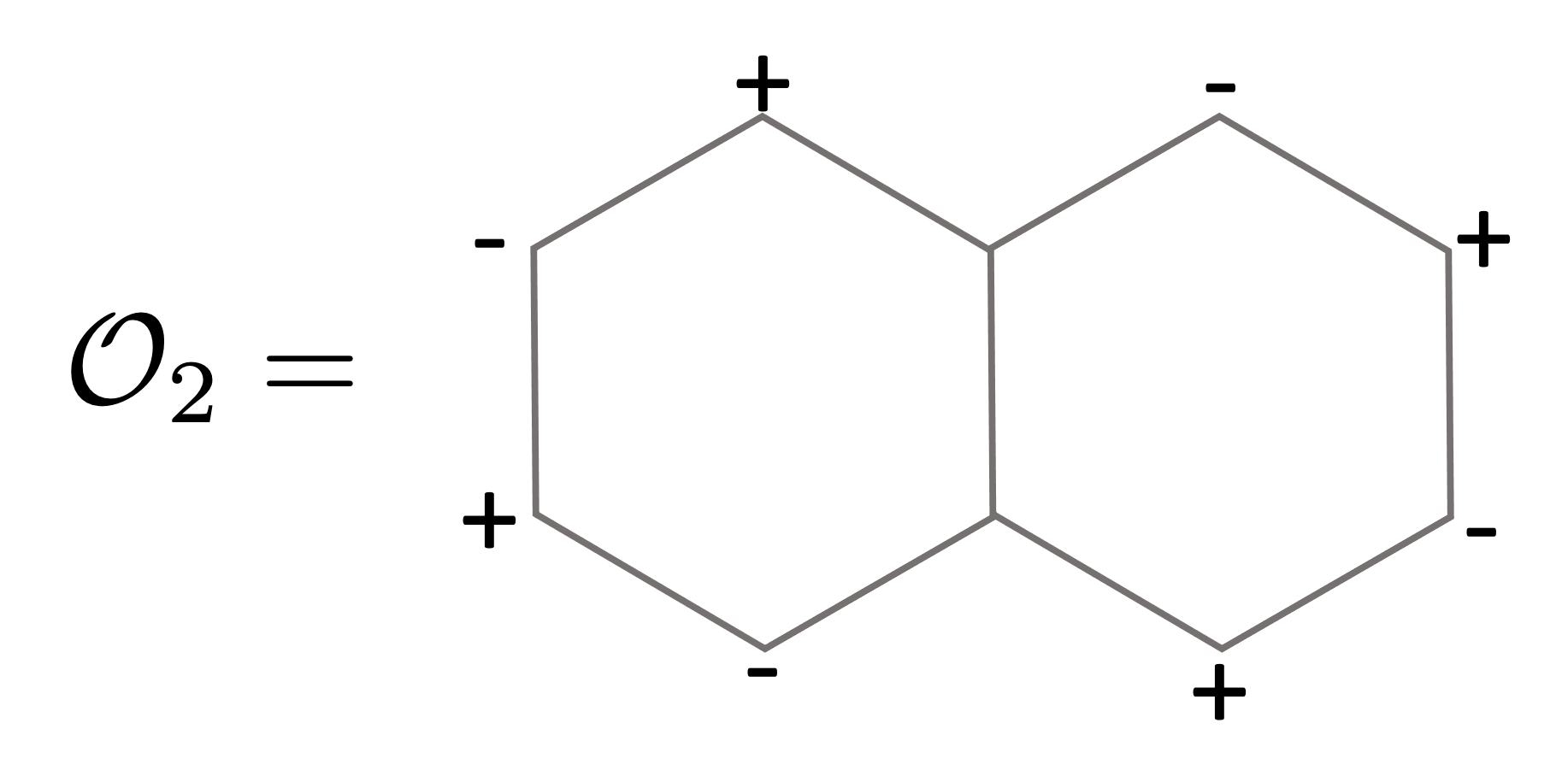}
    \caption{Lowest order ring exchange operators for filling $-10/3$. The $+$($-$) sign denotes the fermion creation(annihilation) operator at the corresponding sites.}
    \label{fig:ring2}
\end{figure}

Now, let us consider the effect of quantum fluctuation. The leading order ring exchange operators can be obtained via a similar procedure as discussed in the main text. We write the lowest order operator as $\mathcal{O}_2$, whose definition is given by Fig.\ref{fig:ring2} and there are also three different orientations related by $C_3$ rotation. Note that among all the four phases we considered previously, $\mathcal{O}_2$ only acts non-trivially for the zig-zag phase and annihilates all the other three phases. Therefore, the zig-zag phase is analogous to columnar order and one can also construct plaquette-like ordering based on $\mathcal{O}_2$. These are the phases that quantum fluctuation would likely favor.

\begin{figure}
    \centering
    \includegraphics[width = 0.45\textwidth]{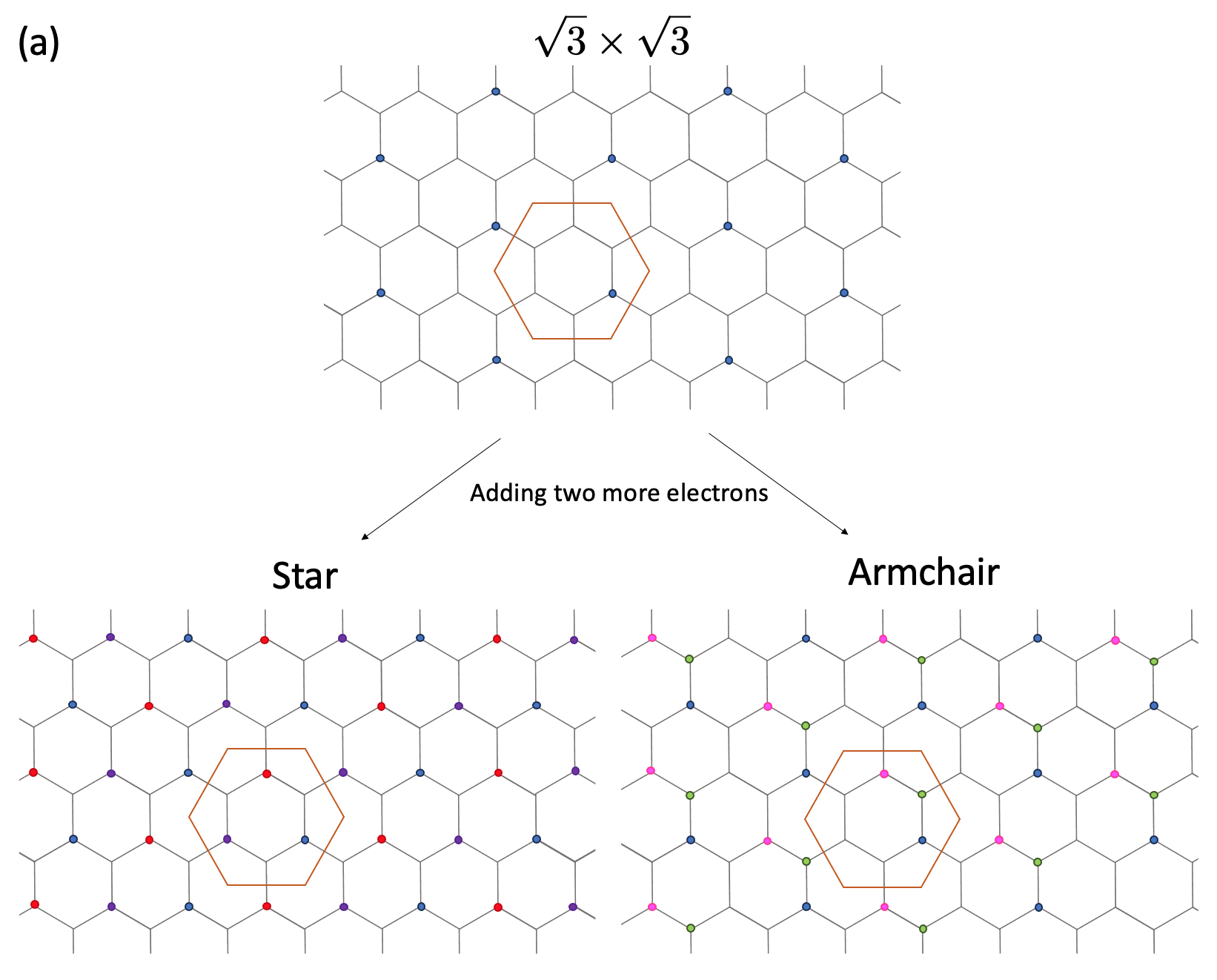}

    \includegraphics[width = 0.45\textwidth]{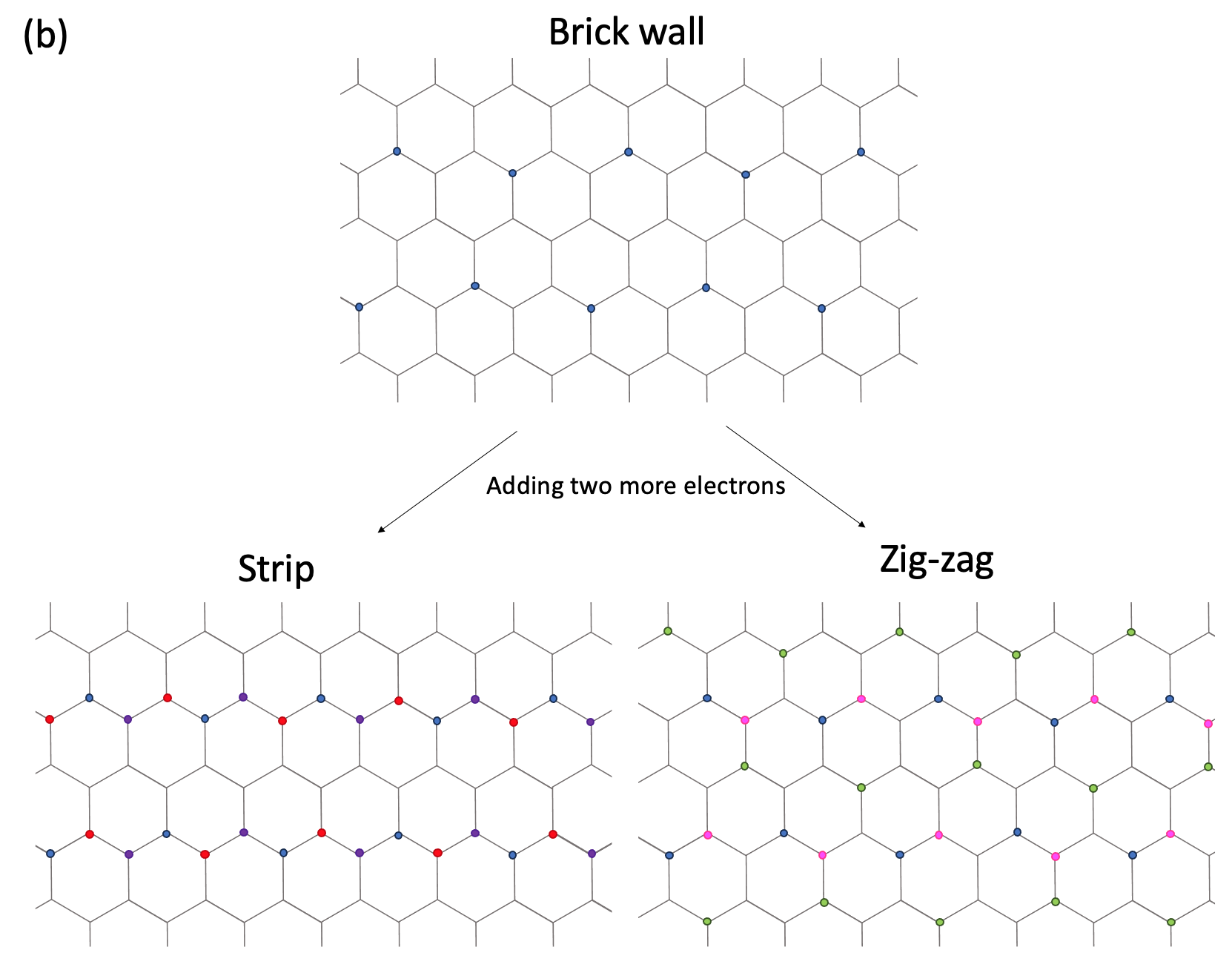}
    \caption{Relationship between the phases at $-11/3$ filling and $-1$ filling. (a) Adding two electrons per plaquette to the $\sqrt{3} \times \sqrt{3}$ phase gives rise to star or armchair phases. The star phase is sublattice polarized. The orange hexagon denotes the unit-cell. (b) Adding two electrons per plaquette to the brick-wall phase gives rise to strip or zig-zag phases. }
    \label{fig:four_phases_3}
\end{figure}

Next, we consider filling $-1$, there are also four different phases in strong coupling limit considering the Hamiltonian in Eq.\ref{eq:H_int_3} (see Fig.\ref{fig:four_phases_3}). And the competition between them is summarized in Table.\ref{tab:energy_four_phases}. The lowest order ring-exchange operator is the $\mathcal{O}$ defined in the main text. It is readily seen that $\mathcal{O}$ only acts non-trivially on the star phase. 

Let us briefly comment on the case with double occupancy. The local Hilbert space of the doubly occupied site is six-dimension, and it forms a 6-d irrep of the $SU(4)$ spin-valley symmetry. Since the $SU(4)$ symmetry is approximate, different spin-valley configurations will split, especially for TBG away from the magic angle. In general, $SU(4)$ breaks into $SU(2) \times SU(2) \times U(1)_v$. Whether the doubly occupied sits favors valley-polarized, spin-singlet or spin-polarized, valley-singlet can be detected by measuring transport in the presence of an in-plane magnetic field. If the spin-triplet state is favored at zero field, the resistance will increase monotonically with the in-plane field due to the Zeeman energy. If the spin-singlet state is favored at zero field, the resistance will have non-monotonic behavior under the field and it will decrease and then increase, due to the competition between the spin-singlet, valley-polarized state, and spin-polarized, valley-singlet state.

\section{H.Twisted trilayer graphene}
Under zero displacement field, the BM model for mirror-symmetric twisted trilayer graphene (MSTG) can be viewed as decoupled twisted bilayer graphene and two Dirac cones (one for each valley)\cite{Khalaf2019Phys.Rev.B}. And the TBG and the Dirac cone degrees of freedom coupled to each other via density-density interaction. Therefore, the Hamiltonian of MSTG can be written as,
\begin{equation}
    \begin{split}
        H_{MSTG} = H_{TBG} + H_{Dirac} + H_{TD},
    \end{split}
\end{equation}
where $H_{TBG}$ and $H_{Dirac}$ include both the kinetic energy and projected interaction, and $H_{TD}$ denotes the interaction between the two. $H_{TBG}$ is the focus of the main text. The analogous correlated states in MSTG at 1/3 fillings are the ones we studied in the main text, with the additional Dirac cones at charge neutrality. Since the density of states of the Dirac cone at neutrality is zero, the interaction term $H_{TD}$ vanishes. And therefore the average energy of the correlated insulator is $E_{MSTG}^0 = E_{TBG}^0 + E_{Dirac}^0 = \langle H_{TBG} \rangle_0 + \langle H_{Dirac} \rangle_0$. 

If we ignore the dispersion of TBG, the competing state would be to hole dope the Dirac cone and populate the flat band of TBG with the same number of electrons. Such a state has lower kinetic energy due to the hole-doped Dirac cone but can have increased interaction energy due to the interaction within TBG.  Let us consider an ansatz wave function for such competing state as $|\psi^\nu \rangle \equiv |\psi_{TBG}^{1/3+\nu}\rangle \otimes |\psi_{Dirac}^\nu \rangle$, where $\nu$ denotes the amount of doped hole into the Dirac cones. And the proposed correlated phase corresponds to $\nu = 0$. The energy density difference between the two states is,
\begin{equation}
    \begin{split}
        \Delta E_{MSTG} &\equiv \frac{1}{N}\langle \psi^\nu | H_{MSTG} |\psi^\nu \rangle - \frac{1}{N}\langle \psi^0 | H_{MSTG} |\psi^0 \rangle\\
        &=- E_{K,Dirac}^\nu + \Delta H_{int,TBG}^\nu  - \tilde{V}_0 (1/3+ \nu) \nu.
    \end{split}
\end{equation}
In the last line of the above equation, the first term is the kinetic energy of the hole Fermi surface. The second term denotes the change in interaction energy of the TBG sector. The third term denotes the attractive interaction between the holes of the Dirac cone and the electrons of the TBG, where $\tilde{V}_0$ is the integral of the projected Coulomb interaction over the whole space. And to get an estimation of the energy scale, we can approximate $ \Delta H_{int,TBG}^\nu$ by a coarse-grained uniform charge distribution and ignore the exchange interaction. Therefore, $ \Delta H_{int,TBG}^\nu \approx \tilde{V}_0 (2/3 + \nu) \nu$ and $\Delta E_{MSTG} \approx \tilde{V}_0 \nu/3 - E_{K,Dirac}^\nu$. Since $E_{K,Dirac}^\nu \sim \nu^{3/2}$, for small $\nu$, the states with $\nu = 0$ always have lower energy. However, for large $\nu$, a hole-doped Dirac cone could be favored, depending on the competition between the repulsive interaction and the kinetic energy of the Dirac cones. Thus, for strong enough repulsive interaction, we expect the MSTG at 1/3 filling can also be described by the model studied in the main text, and host correlated states alike, with additional Dirac cones at charge neutrality.
%}

\pagebreak
\widetext
\bibliographystyle{apsrev4-2}
\bibliography{Moire-fracton}